\documentclass[%
 amsmath,amssymb,
 twocolumn,
]{revtex4-2}

\usepackage{graphicx}
\usepackage{bm}
\usepackage[english]{babel}
\usepackage[T1]{fontenc}
\usepackage{upgreek}
\usepackage{textcomp} 				
\usepackage{multirow}				
\usepackage{amssymb,amsmath}		
\usepackage{siunitx}
\DeclareSIUnit\wn{\cm\tothe{-1}}
\usepackage{verbatim}
\subequations
\usepackage{dblfloatfix} 
\usepackage{soul}

\renewcommand{\paragraph}[1]{{\bf #1}}
\newcommand{\BVO}{BiVO$_4$~}

\begin{document}

\preprint{AIP/123-QED}

\title{Phonon-phonon coupling in bismuth vanadate over a large temperature range across the monoclinic phase}
\author{Christina Hill}
\affiliation{Materials Research and Technology Department, Luxembourg Institute of Science and Technology, 41 rue du Brill, L-4422 Belvaux, Luxembourg}
\affiliation{University of Luxembourg, 41 rue du Brill, L-4422 Belvaux, Luxembourg}
\author{Georgy Gordeev}
\author{Mael Guennou}
\affiliation{University of Luxembourg, 41 rue du Brill, L-4422 Belvaux, Luxembourg}

\date{\today}

\begin{abstract}

In this work we study phonon-phonon coupling in bismuth vanadate (BiVO$_4$), known for its second-order transition involving a variety of coupling mechanisms. Using Raman spectroscopy as a probe, we identify two optical coupled phonon modes of the VO$_4$ tetrahedron and study them by varying light polarization and temperature. The coupling manifests in non-Lorentzian line-shapes of Raman peaks and frequency shifts. We use theoretical framework of coupled damped harmonic oscillators to model the coupling and capture the phenomena in the temperature evolution of the coupling parameters. The coupling is negligible at temperatures below 100 K and later increases in magnitude with temperature until 400 K. The sign of the coupling parameter depends on the light polarization direction, causing either phonon attraction or repulsion. After 400 K the phonon-phonon coupling diminishes when approaching phase transition at which the phonon modes change their symmetry and the coupling is no longer allowed.
\end{abstract}

\maketitle
\section*{Introduction}
Raman spectroscopy is a powerful technique to study lattice dynamics in ferroic oxides. In these materials, the phonon frequency strongly depends on external parameters such as temperature or pressure as well as on the order parameter for example the strain in the material. In order to determine the phonon frequency precisely a proper model to describe the line-shape of the phonon mode should be used. Typically, Raman bands are modelled by a Lorentzian function derived from the classical damped harmonic oscillator model in the approximation of low damping. The eigenfrequency is an important parameter to follow one particular phonon mode across the phase transition. 
In practice, various phenomena may lead to a deviation from a perfect Lorentzian line-shape: Gaussian like broadening due to instrumental effects \cite{Meier2005, chen_automated_2016}, the Doppler broadening in gases \cite{alsmeyer_automatic_2004}, the superimposition of Raman bands as in silicate glass\cite{mysen_1982}, a strong anharmonic potential of atomic bands\cite{gouadec_raman_2007} as in ZnO\cite{cusco_temperature_2007} and in PbTiO$_3$ \cite{schwarz_asymmetric_1997} or sample related effects observed for example in nanoparticales \cite{adu_raman_2006,nemanich_light_1981, korepanov_quantum-chemical_nodate, gouadec_raman_2007}, graphene and related systems \cite{martins_ferreira_evolution_2010, huang_phonon_2009} or in polymers\cite{lehnert_comparative_1997}. Another source for deviations from Lorentzian line-shape are coupling effects of various kind, i.e. when the system cannot be reduced to as set of independent oscillators. Coupling between two phonon bands is discussed in literature for quartz \cite{scott_evidence_1968},BiTO$_3$ \cite{Chaves1974} and AlPO$_4$\cite{scott_hybrid_1970, scott_soft-mode_1974}. In ferroelectric KPD (KH$_2$PO$_4$) \cite{she_effect_1972} and in the arsenates CsH$_2$AsO$_4$ and KH$_2$AsO$_4$ \cite{katiyar_proton-phonon_1971}, proton-phonon coupling causes the deviation from a Lorentzian line-shape. Another widely discussed phenomenon is the coupling between one phonon and a broad continuum \cite{rousseau_auger-like_1968,ager_fano_1995, burke_raman_2010, magidson_fano-type_2002} in literature referred to as Auger-like resonant interference\cite{rousseau_auger-like_1968}, Fano-interference \cite{burke_raman_2010, magidson_fano-type_2002} or Breit-Wigner-Fano coupling \cite{fano_effects_nodate, breit_capture_nodate, brown_origin_2001,burke_raman_2010, Gordeev2016a}. Coupling effects typically cause an asymmetric line-shape. In that case, the eigenfrequency no longer converges with central frequency of the Raman mode. Modeling correctly the Raman line-shape is key to extract the eigenfrequency and therefore identify the Raman mode correctly.

The coupled damped harmonic oscillator (coupled oscillator) model has first been introduced by Barker and Hopfield.\cite{barker_coupled-optical-phonon-mode_1964} They successfully applied it to explain the dielectric properties of various perovskites. Later, Scott\cite{scott_hybrid_1970}, Zawadowski and Ruvalds \cite{ruvalds_indirect_1970}, used the Green's formalism to derive a more generalized form to model phonon-phonon coupling in AlPO$_4$. Katiyar et al. \cite{katiyar_proton-phonon_1971} and She et al. \cite{she_effect_1972} then applied the Greens formalism for proton-phonon coupling in KH$_2$AsO$_4$ (KDA) and KH$_2$PO$_4$ (KDP), respectively. The coupling between phonon modes is a temperature induced effect and \citet{Chaves1974} showed that only modes of same symmetry can couple. While the coupled oscillators model is known, its implementation can be challenging and misleading. In many cases, multiple solutions provide a good fit to the data leading to very different physical conclusions. The implementation of physical meaningful constraints on the modeling parameters is a key to overcome this problem. In the literature, constraining the complex coupling parameter assuming either pure real \cite{katiyar_proton-phonon_1971,Chaves1974,she_effect_1972, scalabrin_temperature_1977} or pure imaginary  \cite{lagakos_preliminary_1974,bozinis_optical_1976} coupling has been widely used. However, the physical correctness of the coupling parameters needs to be carefully revised. This can be done by fitting the line-shape of data measured at many temperatures (or pressures or electric fields). However there are not many examples in the literature where this has been done.
\BVO is a nice example to study effects of phonon-phonon coupling since it is well known for the phase transition driven by a B$_{g}$ optical soft mode (tetragonal) coupled to an acoustic mode.\cite{benyuan_soft_1981, avakyants_inelastic_2004} The phase transition in \BVO is classified as 2$^{nd}$ order from high temperature tetragonal phase (space group $I4_1/a$) to a low temperature monoclinic phase (space group $C27c$, most commonly described in non-standard setting $I2/a$), also referred to as Scheelite and Fergusonite structures respectively. The Raman active modes are known from standard symmetry analysis and they have been successfully assigned.\cite{avakyants_inelastic_2004, pellicer-porres_phase_2018} Their frequency shifts with temperature and pressure across the phase transition have been reported.\cite{pellicer-porres_phase_2018} However, we observed line-shape asymmetries of the vibration modes of the VO$_4$ tetrahedra that strongly depends on temperature and on the polarization conditions of the experiment. 

In this work we measured the Raman modes corresponding to the vibration modes of the VO$_4$ tetrahedron in \BVO over large temperature range at different light polarization geometries. We observed anomalies in the line-shape and in the frequency position of these Raman modes that suggest coupling between phonons of the same symmetry. We applied the coupled damped harmonic oscillators model to describe the observed phonon-phonon coupling for two different polarization conditions. Finally We studied the phonon-phonon coupling in a high temperature range from 10 K to 510 K across the phase transition.

\section*{Experimental Methods}
We investigated Raman spectra of \BVO single crystals over a large temperature range. The Raman spectrometer InVia from Renishaw in backscattering configuration with a microscope, 50x objective was used. For all measurements, we used \SI{633}{\nm} (\SI{1.96}{\eV}) laser excitation. We use the Porto's notation i(jk)l to label the different polarization geometries, where i and l correspond to the propagation direction and j and k the polarization direction with respect to the crystallographic axis. Linkam temperature cell THMS600 was used to cover temperatures from \SI{80}{\K} to \SI{500}{\K}. The helium flow cryostat Microstat from Oxford Instruments was used to cover the low temperature range from \SI{10}{\K} to \SI{300}{\K}. 

\begin{figure}[htp]
    \centering
    \includegraphics[width=0.8\linewidth]{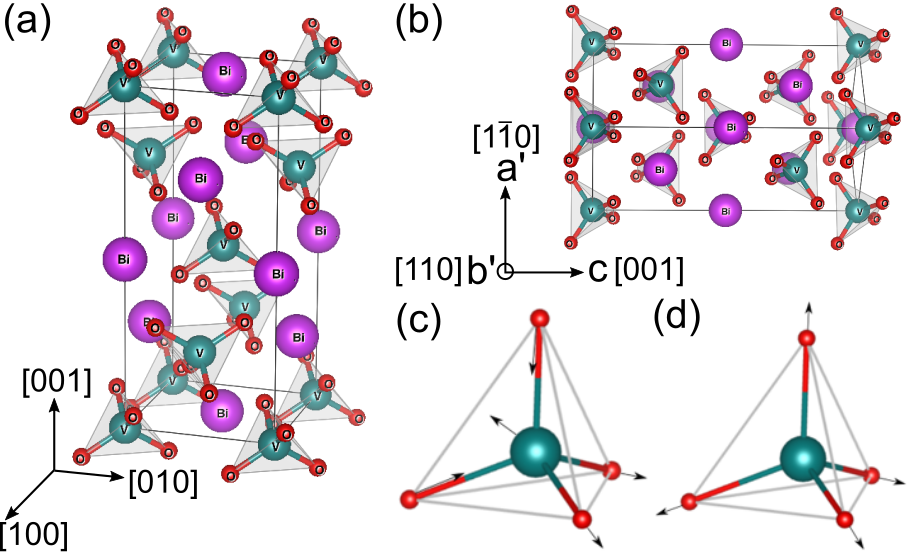}
    \caption{Schematic representation of (a) the tetragonal unit cell of bismuth vanadate and (b) view on the $(001)$ surface. Visualization of the symmetries of the atomic vibrations of the oxygen tetrahedron corresponding to (c) the  $A_g^7$ and (d) the $A_g^8$.}
    \label{fig:crystal}
\end{figure}

The crystal structure of \BVO is tetragonal at high temperature with a phase transition at \SI{523}{\K} to monoclinic. Fig.~\ref{fig:crystal}~(a) shows the tetragonal unit cell with 4-fold rotational axis along $[001]$. In the low-temperature monoclinic phase, the 4-fold rotational axis becomes a 2-fold rotational axis, the so called principle axis. For this study we used a single with the surface orientation $(110)$. The view on the $(110)$ surface is given in Fig.~\ref{fig:crystal}~(b) in which the principal axis is labeled with $c$ and the $[1\bar{1}0]$-axis with $b'$. The direction $[110]$ pointing out of the surface plane is labeled with $a'$. From standard symmetry analysis, we expect 8 phonon modes of $A_g$ symmetry and 10 phonon modes of $B_g$ symmetry in the monoclinic phase.\cite{pellicer-porres_phase_2018, avakyants_inelastic_2004}

The Raman (Stokes) scattering intensity is correlated to the imaginary part of the susceptibility $\chi''$
\begin{equation} 
S(\omega) \propto \chi''(\omega) [n(\omega)+1],
\end{equation}

\noindent where $n$ is the phonon population factor given by the Bose thermal factor $n(\omega)=(e^{\hbar\omega/kT}-1)^{-1}$  and T is the absolute temperature of the sample. All presented data have been divided by the thermal Bose factor ${n(\omega)+1}$ so that they are directly proportional to the imaginary part of the susceptibility. The description of the Raman scattering intensity strongly depends on the model chosen for the susceptibility.

\section*{Theoretical Methods}
\begin{figure}[htp]
    \centering
    \includegraphics[width=0.6\linewidth]{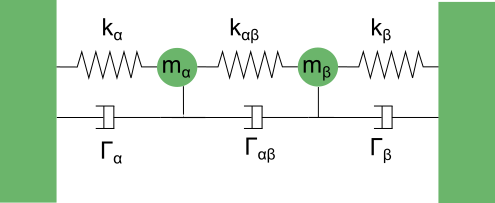}
    \caption{Mechanical model of two damped harmonic oscillators $\alpha$ and $\beta$ described by spring constants $k_\alpha$, $k_\beta$ and the damping coefficients $\Gamma_\alpha$,$\Gamma_\beta$. The coupling is complex and given by the coupling spring constant $k_{\alpha\beta}$ and the damping coefficient $\Gamma_{\alpha\beta}$.}
    \label{fig:Greens}
\end{figure}

In the following the coupled damped harmonic oscillators (coupled oscillators) model is explained. The basic concept has been developed by Barker and Hopfield.\cite{barker_coupled-optical-phonon-mode_1964} Fig.~\ref{fig:Greens} shows the mechanical model for a system of two coupled phonons. Each phonon is described as a damped harmonic oscillator with an effective mass $m$, spring constant $k$ and damping coefficient $\Gamma$. The two masses $\alpha$ and $\beta$ are mechanically connected by an additional spring $k_{\alpha\beta}$ and dashpot $\Gamma_{\alpha\beta}$.

The Green's function may be written as $G^{-1}=g=-\omega^2 \boldsymbol{I_2} +i\omega \boldsymbol{\Gamma} + \boldsymbol{K}$ in which $\boldsymbol{I_2}$ being the unit matrix and $\boldsymbol{K}$ and $\boldsymbol{\Gamma}$ being restoring force and damping matrices. Using the Green's function formalism a system of two coupled damped harmonic oscillators\cite{katiyar_proton-phonon_1971,she_effect_1972} is defined by
\begin{equation} \label{eq:Greens}
\begin{bmatrix}
\omega_\alpha^2-\omega^2+i\omega\Gamma_\alpha & \Delta^2+i\omega\Gamma_{\alpha\beta}\\
\Delta^2+i\omega\Gamma_{\alpha\beta} & \omega_\beta^2-\omega^2+i\omega\Gamma_\beta
\end{bmatrix}
\begin{bmatrix}
G_{\alpha\alpha} & G_{\alpha\beta}\\
G_{\alpha\beta} & G_{\beta\beta}
\end{bmatrix} =
\begin{bmatrix}
1 & 0\\
0 & 1
\end{bmatrix}\end{equation}.

 \noindent The parameters $\Delta$, $\Gamma_\alpha$, $\Gamma_\beta$ and $\Gamma_{\alpha\beta}$ are frequency independent. The effective mass $m$ is set to unity. The coupling is described by $\Delta$ and $\Gamma_{\alpha\beta}$ which is the force constant of the spring and the damping coefficient of the dashpot connecting the two oscillators. $\Delta$ and $\Gamma_{\alpha\beta}$ must be constraint by $det(K),det(\Gamma) \ge 0$.\cite{takagi_coupled_1983} This means that the coupling cannot be arbitrary large. The frequencies $\omega_\alpha$ and $\omega_\beta$ correspond to the eigenfrequencies of the uncoupled phonons. The eigenfrequency squared  is equal to the spring constant divided by the effective mass $m$, $\omega_{\alpha/\beta}^2=k_{\alpha/\beta}/m_{\alpha/\beta}$. $\Gamma_\alpha$ and $\Gamma_\beta$ are the damping coefficient of the uncoupled vibration modes $\alpha$ and $\beta$.

Equation~\ref{eq:Greens} is solved to give the following Green's functions coefficients:
\begin{equation}
    G_{\alpha\alpha}=(\omega_\beta^2-\omega^2+i\omega\Gamma_\beta)/D
\end{equation}
\begin{equation}
    G_{\alpha\beta}=-(\Delta^2+i\omega\Gamma_{\alpha\beta})/D
\end{equation}
\begin{equation}
    G_{\beta\beta}=(\omega_\alpha^2-\omega^2+i\omega\Gamma_\alpha)/D
\end{equation}
where $D=(\omega_\alpha^2-\omega^2+i\omega\Gamma_\alpha)(\omega_\beta^2-\omega^2+i\omega\Gamma_\beta)-(\Delta^2+i\omega\Gamma_{\alpha\beta})^2$. 

Equation~\ref{eq:Greens} allows arbitrary choice of diagonalization and can produce an infinite number of solutions.\cite{lowndes_ferroelectric_1974} Some physical assumptions have to be made to filter reasonable solutions. Many authors\cite{mazzacurati_indirect_1976,takagi_coupled_1983} proposed to make the assumptions on the complex coupling of modes in which the two limiting cases occur for only real coupling ($\Gamma_{\alpha\beta}=0$) and only imaginary coupling ($\Delta=0$). The parameters obtained in the two cases are entirely different as well as the physical interpretation, see supplementary information. In this work, we suggest to make use of the temperature dependent data as well as the polarization dependent data to find an unique solution to the coupled oscillators model.

Using matrix notation, the complex susceptibility $\chi(\omega)$ for coupled oscillators model is given by $\chi(\omega)=\tilde{P} \cdot G\cdot P$ in which $P$ and $\tilde{P}$ are one column and one row vectors. The imaginary part of the complex susceptibility $\chi''(\omega)$ is then given by
\begin{equation}
\chi''(\omega) = -Im[P_\alpha^2G_{\alpha\alpha}(\omega)+2P_\alpha P_\beta G_{\alpha\beta}(\omega)+P_\beta^2 G_{\beta\beta}(\omega)],
\end{equation}

\noindent $P_\alpha$, $P_\beta$ being the oscillator strengths of the vibration mode $\alpha$ and $\beta$ and $\chi''$ is proportional to the  intensity of the Raman spectrum.

\section*{Experimental results}

First, we discuss the evidences for coupling phenomena in the high-frequency Raman spectrum of BiVO$_4$. Raman spectra showing all 8 Raman modes of $A_g$ symmetry are shown in the supplementary information. For most of these Raman modes, the eigenfrequency does not dependent on the polarization conditions chosen in the experiment, however this is different for the high-frequency $A_g^7$ mode that shows coupling signatures with the $A_g^8$ mode. Fig.~\ref{fig:crystal}~(b) shows $A_g^7$ near \SI{705}{\wn} and $A_g^8$ at \SI{830}{\wn} measured at room temperature. The symmetries of the atomic vibrations corresponding to the polar $A_g^7$ and the non-polar $A_g^8$ mode are shown in Figs.~\ref{fig:crystal}~(c) and (d), respectively. 

\begin{figure}[htp]
    \centering
    \includegraphics[width=\linewidth]{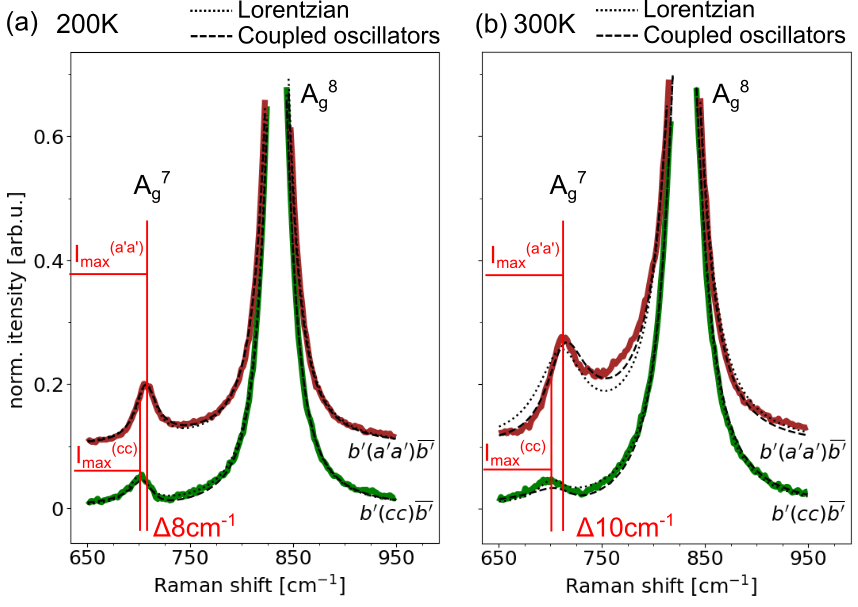}
    \caption{Raman spectra of the high-frequency $A_g^7$ and $A_g^8$ modes measured at different polarization geometries $b'(cc)\bar{b}'$ and $b'(a'a')\bar{b}'$ at (a) \SI{200}{\K} and b) \SI{300}{\K} with a \SI{1.96}{\eV} excitation. Dashed lines show the fit assuming two Lorentzian oscillators in black and coupled oscillators model in blue. The fitting parameters are given in Table~\ref{tab:Tab1}.}
    \label{fig:DHOvsCDHO_RT}
\end{figure}

Fig.~\ref{fig:DHOvsCDHO_RT} compares the phonon modes $A_g^7$ and $A_g^8$ measured at (a) \SI{200}{\K} and (b) room temperature for two polarization geometries. The first observation we made on the raw data is that the central frequency of the $A_g^7$ depends on the polarization geometry, $\Delta \omega=8$ and \SI{10}{\wn} for 200 and 300 K, respectively. Second, at \SI{300}{\K} and for the polarization $b'(a'a')\bar{b}'$ (brown spectrum in Fig.\ref{fig:DHOvsCDHO_RT}~(b)), we see additional intensity in between the two Raman modes.

Lorentz model fails to reproduce the experimental data, see dotted line. At \SI{200}{\K} the Lorenz model only partly converges with the data by assuming non physical behavior: different $A_g^7$ eigenfrequency for different polarization conditions. The fitting parameters are given in Table~\ref{tab:Tab1}. At \SI{300}{\K} the two Lorentzian peaks fit even fails to identify the $A_g^7$ phonon mode, if the frequency is not fixed. There are two reasons for that; first the line-shape of the $A_g^7$ is more asymmetric at higher temperatures and there is significant intensity background in between the two modes. Overall it is clear that the Lorentz model cannot be used for the Raman interpretation since it delivers falsified results.
While the eigenfrequency of the $A_g^8$ mode almost matches for both polarization geometries, the eigenfrequency of the $A_g^7$ mode is  different from one polarization geometry to the other by almost 1.2 $\%$. This is rather unexpected, since mode frequency should be independent upon polarization and can only be explained by the coupled oscillator model.

Now we will discuss the fits using the coupled oscillators model. For the coupled oscillators model we imposed the same eigenfrequency and the damping coefficient for both polarization geometries. Further, we assumed the real part of the coupling parameter $\Delta$ to be zero. For the $A_g^8$ phonon mode, there is little difference between the classical Lorentzian  and the coupled oscillators models. The overall line-shape is well described by both models and the frequency position as well as the damping coefficient of the $A_g^8$ mode is nearly identical for the both models. The major improvement is found for the $A_g^7$ mode at \SI{300}{\K}. The coupled oscillators model succeeds to describe the asymmetry in the line-shape of the $A_g^7$ mode as well as the increased background intensity in between the two modes. This is especially true for the $b'(a'a')\bar{b}'$ polarization geometry. The coupling coefficients are listed in Table \ref{tab:Tab1}, they vary from one polarization geometry to the other. For $b'(cc)\bar{b}'$, the imaginary part of the coupling parameter $\Gamma_{78}$ is positive which correspond to a repulsion of the two phonon modes involved in the coupling. In contrast, for $b'(a'a')\bar{b}'$, the imaginary part $\Gamma_{78}$ is negative leading to an attraction of the two phonon modes.

\begin{table}[h!]
\caption{Comparison of fitting parameters between the classical Lorentzian model and coupled oscillators model at \SI{200}{\K} and \SI{300}{\K}. Here, the real part of the coupling $\Delta$ is fixed to zero. The eigenfrequencies $\omega$ and damping coefficients $\Gamma$ are given in units of \SI{}{\wn}.}
\begin{center}
\begin{tabular}{c| c c | c  c }
\hline
& \multicolumn{2}{ c |}{\textbf{200K}} & \multicolumn{2}{ c }{\textbf{300K}}\\ 
& $b'(a'a')\bar{b}'$ & $b'(cc)\bar{b}'$ & $b'(a'a')\bar{b}'$ & $b'(cc)\bar{b}'$ \\ 
\hline 
\multicolumn{5}{ l  }{\textbf{Lorentzian model}}\\
\hline
$\omega_{7}$ & 708 & 700 & 711 (fixed) & 701 (fixed) \\ 
$\Gamma_{7}$ & 24 & 22 & 54 & 31 \\ 
$\omega_{8}$ & 834 & 835 & 830 & 830 \\ 
$\Gamma_{8}$ & 26 & 26 & 36 & 34 \\ \cline{3-5}
\hline \hline
\multicolumn{5}{ l  }{\textbf{Coupled oscillators model}}\\
\hline
$\omega_{7}$ &\multicolumn{2}{ c |}{706} & \multicolumn{2}{ c }{712}\\ 
$\Gamma_{7}$ &\multicolumn{2}{ c |}{25.5} & \multicolumn{2}{c }{47.0}\\ 
$\omega_{8}$ &\multicolumn{2}{ c |}{835} & \multicolumn{2}{c }{831}\\ 
$\Gamma_{8}$ &\multicolumn{2}{c |}{26.2} & \multicolumn{2}{c }{34.9}\\ 
$\Gamma_{78}$ & -5.3 & 12.3 & -15.0 & 22.4 \\ 
\hline
\end{tabular}
\end{center}
\label{tab:Tab1}
\end{table}

\begin{figure}[htp]
    \centering
    \includegraphics[width=\linewidth]{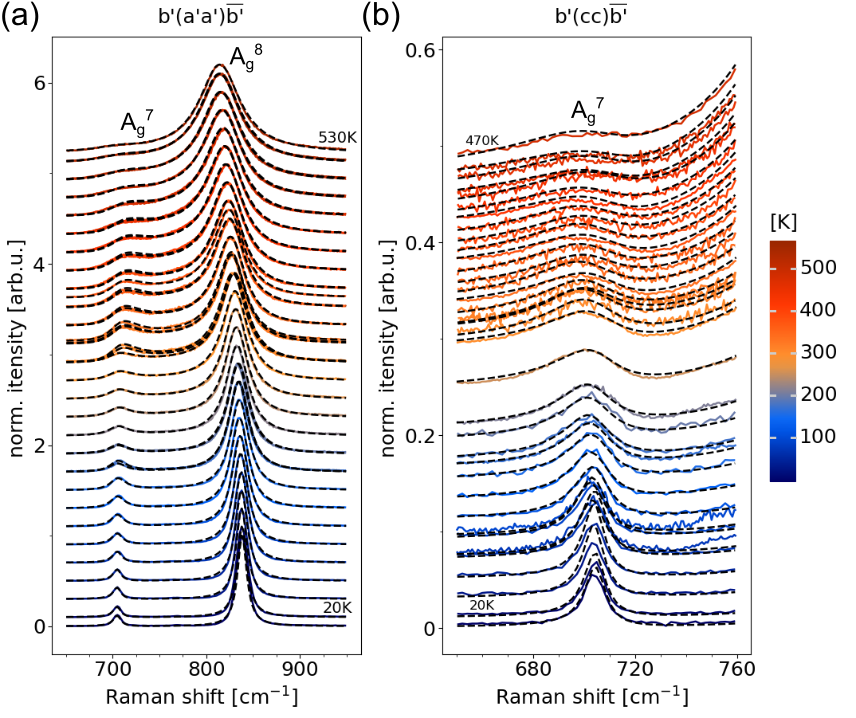}
    \caption{Coupled oscillators model applied to the Raman modes $A_g^7$ and $A_g^8$ at different temperatures for the (a) polarization geometry $b'(a'a')\bar{b}'$ and (b) $b'(cc)\bar{b}'$ . Here, the real part of the coupling $\Delta$ is assumed to be zero. The damping of the individual oscillators $\Gamma_7$ and $\Gamma_8$ is assumed to increase linear with temperature. Only for polarization geometry $b'(a'a')\bar{b}'$, eigenfrequency $\omega_7$ of the $A_g^7$ mode is assumed to be constant.}
    \label{fig:CDHO_HighFreq_T}
\end{figure}

We further study the temperature dependence of phonon-phonon more systematically. Fig.~\ref{fig:CDHO_HighFreq_T}~(a) shows the $A_g^7$ and the $A_g^8$ mode measured at temperatures ranging from \SI{10}{\K} up to phase transition at \SI{523}{\K} for the polarization geometries $b'(a'a')\bar{b}'$. By increasing the temperature we observe typical effects: the modes shift and broadening with temperature. The $A_g^8$ mode clearly shifts to lower frequencies with increasing temperatures whereas the $A_g^7$ seemingly shift to higher frequencies. Also we can directly see the evolution of phonon-phonon coupling. At low temperatures, the two Raman modes are well isolated whereas at higher temperature additional intensity develops in between the modes. The additional intensity increases with increasing temperatures with respect to the intensity contribution from the individual modes. That is why, we expect a strong temperature dependence of the coupling parameter.

Fig.~\ref{fig:CDHO_HighFreq_T}~(b) shows a zoomed spectral range with the $A_g^7$ mode measured with $b'(cc)\bar{b}'$ polarization. In this light polarization the $A_g^7$ mode becomes extremely weak compared to the $A_g^8$ mode while approaching the phase transitions. At temperatures above \SI{470}{\K}, the $A_g^7$ mode is no longer visible in the raw data, see also the supplementary information. Surprisingly, the $A_g^7$ mode shifts to lower frequencies with increasing temperatures which contradicts the measurement for $b'(a'a')\bar{b}'$ shown in Fig.~\ref{fig:CDHO_HighFreq_T}~(a). A phonon-phonon coupling model is required to explain this unconventional behavior. 

The black dashed lines in Fig.~\ref{fig:CDHO_HighFreq_T} are fits based on the coupled oscillators model. For both polarization geometries we find a good agreement between the model and experiment. As mentioned in the methods section, the coupled oscillators model is overparameterized so that assumptions on the fitting parameters have to be made. We assumed i) the coupling parameter $\Delta$ to be zero, ii) the individual damping $\Gamma_7$, $\Gamma_8$ to increase linearly with temperature at a constant slope of \SI{0.1}{} and \SI{0.08}{\wn \per \K}, respectively as expected for two-phonon decay and iii) the eigenfrequency of the $A_g^7$ mode to be constant as a function of temperature for the light polarization $b'(a'a')\bar{b}'$. As a final element we implemented a $weight$ equal to the inverse of the intensity in the minimization algorithm. The implementation of a $weight$ compensates for the huge intensity difference between the $A_g^7$ and the $A_g^8$ mode and is introduced as a multiplication factor so that $weight \cdot (data-fit)^2$ is minimized . This non-standard implementation in the minimization algorithm was key to find fitting solutions for the data measured at $b'(cc)\bar{b}'$ polarization and at high temperatures. The intensity difference is not as drastic for the $b'(a'a')\bar{b}'$ light polarization, so that in this case the $weight$ was set to one.  

\begin{figure}[htp]
    \centering
    \includegraphics[width=\linewidth]{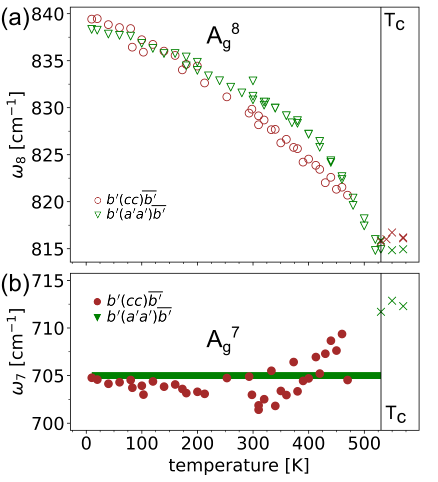}
    \caption{Fitting parameters of the coupled oscillators model applied to the Raman modes $A_g^7$ and $A_g^8$ as a function of temperature: eigenfrequency for the two polarization geometries of (a) the low intensity $A_g^7$ mode and (b) of the $A_g^8$ mode.}
    \label{fig:Eig_Freq}
\end{figure}

Fig.~\ref{fig:Eig_Freq}~(a) shows the eigenfrequency $\omega_8$ of the $A_g^8$ mode extracted from the fits shown in Fig.\ref{fig:CDHO_HighFreq_T} as a function of temperature for the two polarization geometries. The data points above the phase transition are included and symbolized with a cross. The eigenfrequency $\omega_8$ strongly decreases with increasing temperature as it is visible in the raw data shown in Fig.\ref{fig:CDHO_HighFreq_T}. Across the phase transition at $T_c = 510 K$, the change in the eigenfrequency is continuous. This is expected for the second-order phase transition, but the slope clearly changes at the phase transition. 

Fig.~\ref{fig:Eig_Freq}~(b) shows the eigenfrequency $\omega_7$ of the $A_g^7$ as a function of temperature for the two polarization geometries. For the polarization geometry $b'(cc)\bar{b}'$, the eigenfrequency of the $A_g^7$ is constant over a large temperature range, before the values slightly increases while approaching the phase transition. A constant temperature trend of eigenfrequency $\omega_7$ is only obtained by the coupled oscillators model, see also Lorentzian fitting as a function of temperature in the supplementary information. Even though we do not have data at temperatures above \SI{470}{\K}, the change in the eigenfrequency seems to be continuous similar to the discussion for the eigenfrequency of the $A_g^8$ mode.

To find a reliable temperature trend for the coupling parameter we assumed the eigenfrequency $\omega_7$ to be independent on temperature for the second polarization geometry $b'(a'a')\bar{b}'$. The intensity of the mode is relatively low therefore constraining it's frequency substantially improved the reliability of the fit. This is fully consistent with lattice dynamics where in non-polar crystals the $\omega_7$ should be polarization independent.  Therefore the  $\omega_7$ was set to \SI{705}{\wn}, same as the average frequency found in the $b'(cc)\bar{b}'$ polarization direction. While the eigenfrequency of the $A_g^8$ mode is robust against different modeling approaches, coupled oscillators model succeeds to deliver physical results for the $A_g^7$ mode, where experimental data are reproduced with the same eigenfrequency for both polarization directions in the entire measured temperature range.

\begin{figure}[htp]
    \centering
    \includegraphics[width=\linewidth]{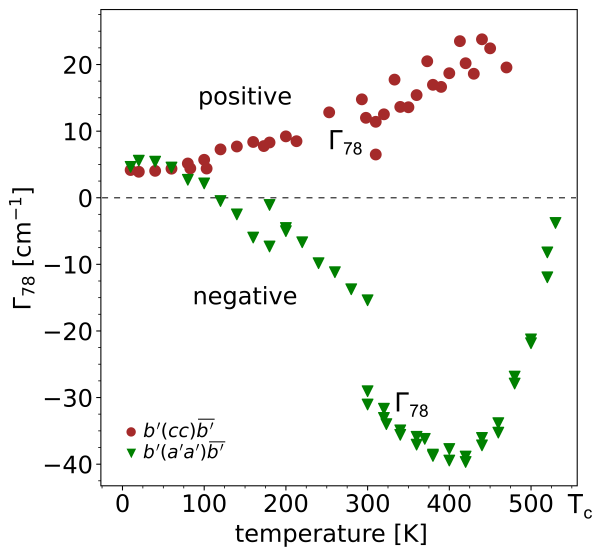}
    \caption{Coupling parameter $\Gamma_{78}$ as a function of temperature for the polarization geometry $b'(cc)\bar{b'}$ and $b'(a'a')\bar{b'}$. Because the intensity of the $A_g^7$ mode diminishes while approaching the phase transition, there are no data for $\Gamma_{78}$ at temperatures above \SI{470}{\K} in case of the $b'(cc)\bar{b'}$ polarization.}
    \label{fig:Fit_params_CDHO}
\end{figure}

Fig.~\ref{fig:Fit_params_CDHO} shows the coupling parameter $\Gamma_{78}$ as a function of temperature for the polarization geometries $b'(cc)\bar{b}'$ and $b'(a'a')\bar{b}'$, respectively. In the polarization geometry $b'(cc)\bar{b}'$, the coupling parameter is positive and increases continuously with increasing temperatures below 400 K, see red circles in Figure ~\ref{fig:Fit_params_CDHO}. On the other hand the $\Gamma_{78}$ for $b'(cc)\bar{b}'$ has an opposite trend in the same temperature range. The overall magnitude of the coupling increases, but the sign is negative, see green triangles in Figure ~\ref{fig:Fit_params_CDHO}. The data between \SI{160}{\K} and \SI{300}{\K} have been measured at a later experiment and show a small offset to overall temperature trend. Overall we have a precise control over the phonon-phonon coupling, where the coupling strength is set by temperature and the coupling sign by the light polarization. 




Now we discus the coupling behavior near the phase transition. When going from monoclic to tetragonal the phonon symmetries change, along with the selection rules. The $A_g^7$ changes its symmetry to a $B_g$ symmetry while the $A_g^8$ remains of $A_g$ symmetry. As a consequence, we expect  no coupling between these two phonons in the high-temperature tetragonal phase.  This is reflected in change of the temperature trend found the coupling parameter for the light polarization $b'(a'a')\bar{b}'$ in which the coupling parameter decreases while approaching the phase transition.\par

The situation is slightly different for the other light polarization $b'(cc)\bar{b}'$, where only modes of $A_g$ symmetry are allowed in the tetragonal phase. \cite{pellicer-porres_phase_2018} Therefore, the Raman intensity of the $A_g^7$ mode gradually decreases with temperature until it disappears in the tetragonal phase. This is nicely observed in the waterfall plot shown in Fig~\ref{fig:CDHO_HighFreq_T}~(b). The ratio of the relative intensities is plotted in the supplementary information. Therefore the symbols for $\Gamma_{78}$ at $b'(cc)\bar{b}'$ are missing in Figure  ~\ref{fig:Fit_params_CDHO}. Overall we in \BVO we find a beautiful demonstration how the change of phonon symmetries influences phonon-phonon coupling. We confirm that only phonon modes of the same symmetry can couple as predicted by \citet{Chaves1974}. 

\section*{Conclusion}
In \BVO we studied coupling between two optical phonons with the $A_g$ symmetry. We show that the phonon-phonon coupling causes the asymmetric line-shapes of the $A_g^7$ and $A_g^8$ mode at temperatures above 100 K.  Depending on the polarization geometry, the sign of coupling parameter is either positive or negative referring to either repulsion or attraction of phonon modes. The coupling strength, expressed in the absolute value of the coupling parameter, increases with increasing temperatures. At temperatures above \SI{400}{K}, the coupling strength decreases for the $b'(a'a')\bar{b}'$ polarization geometry. This is explained by the change in symmetry of the $A_g^7$ mode, which leads to decoupling of the two phonon modes above the phase transition temperature.

\section*{Acknowledgments}
CH acknowledges funding from the Fond National de la Recherche under Project PRIDE/15/10935404. This  research  was  funded  in  whole, or  in  part,  by  the  Luxembourg  National  Research Fund (FNR), grant reference [INTER/MERA20/14995416/SWIPE]. For the purpose of open access, and in fulfilment of the obligations arising from the grant agreement, the author has applied a Creative  Commons Attribution  4.0  International  (CC  BY  4.0) license  to  any  Author Accepted Manuscript version arising from this submission.

\nocite{*}
\bibliography{ModeCoupG}

\begin{thebibliography}{50}%
\makeatletter
\providecommand \@ifxundefined [1]{%
 \@ifx{#1\undefined}
}%
\providecommand \@ifnum [1]{%
 \ifnum #1\expandafter \@firstoftwo
 \else \expandafter \@secondoftwo
 \fi
}%
\providecommand \@ifx [1]{%
 \ifx #1\expandafter \@firstoftwo
 \else \expandafter \@secondoftwo
 \fi
}%
\providecommand \natexlab [1]{#1}%
\providecommand \enquote  [1]{``#1''}%
\providecommand \bibnamefont  [1]{#1}%
\providecommand \bibfnamefont [1]{#1}%
\providecommand \citenamefont [1]{#1}%
\providecommand \href@noop [0]{\@secondoftwo}%
\providecommand \href [0]{\begingroup \@sanitize@url \@href}%
\providecommand \@href[1]{\@@startlink{#1}\@@href}%
\providecommand \@@href[1]{\endgroup#1\@@endlink}%
\providecommand \@sanitize@url [0]{\catcode `\\12\catcode `\$12\catcode
  `\&12\catcode `\#12\catcode `\^12\catcode `\_12\catcode `\%12\relax}%
\providecommand \@@startlink[1]{}%
\providecommand \@@endlink[0]{}%
\providecommand \url  [0]{\begingroup\@sanitize@url \@url }%
\providecommand \@url [1]{\endgroup\@href {#1}{\urlprefix }}%
\providecommand \urlprefix  [0]{URL }%
\providecommand \Eprint [0]{\href }%
\providecommand \doibase [0]{https://doi.org/}%
\providecommand \selectlanguage [0]{\@gobble}%
\providecommand \bibinfo  [0]{\@secondoftwo}%
\providecommand \bibfield  [0]{\@secondoftwo}%
\providecommand \translation [1]{[#1]}%
\providecommand \BibitemOpen [0]{}%
\providecommand \bibitemStop [0]{}%
\providecommand \bibitemNoStop [0]{.\EOS\space}%
\providecommand \EOS [0]{\spacefactor3000\relax}%
\providecommand \BibitemShut  [1]{\csname bibitem#1\endcsname}%
\let\auto@bib@innerbib\@empty
\bibitem [{\citenamefont {Meier}(2005{\natexlab{a}})}]{Meier2005}%
  \BibitemOpen
  \bibfield  {author} {\bibinfo {author} {\bibfnamefont {R.~J.}\ \bibnamefont
  {Meier}},\ }\bibfield  {title} {\bibinfo {title} {{On art and science in
  curve-fitting vibrational spectra}},\ }\href
  {https://doi.org/10.1016/j.vibspec.2005.03.003} {\bibfield  {journal}
  {\bibinfo  {journal} {Vibrational Spectroscopy}\ }\textbf {\bibinfo {volume}
  {39}},\ \bibinfo {pages} {266} (\bibinfo {year}
  {2005}{\natexlab{a}})}\BibitemShut {NoStop}%
\bibitem [{\citenamefont {Chen}\ and\ \citenamefont
  {Dai}(2016)}]{chen_automated_2016}%
  \BibitemOpen
  \bibfield  {author} {\bibinfo {author} {\bibfnamefont {Y.}~\bibnamefont
  {Chen}}\ and\ \bibinfo {author} {\bibfnamefont {L.}~\bibnamefont {Dai}},\
  }\bibfield  {title} {\bibinfo {title} {{Automated decomposition algorithm for
  Raman spectra based on a Voigt line profile model}},\ }\href
  {https://doi.org/10.1364/ao.55.004085} {\bibfield  {journal} {\bibinfo
  {journal} {Applied Optics}\ }\textbf {\bibinfo {volume} {55}},\ \bibinfo
  {pages} {4085} (\bibinfo {year} {2016})}\BibitemShut {NoStop}%
\bibitem [{\citenamefont {Alsmeyer}\ and\ \citenamefont
  {Marquardt}(2004)}]{alsmeyer_automatic_2004}%
  \BibitemOpen
  \bibfield  {author} {\bibinfo {author} {\bibfnamefont {F.}~\bibnamefont
  {Alsmeyer}}\ and\ \bibinfo {author} {\bibfnamefont {W.}~\bibnamefont
  {Marquardt}},\ }\bibfield  {title} {\bibinfo {title} {{Automatic generation
  of peak-shaped models}},\ }\href {https://doi.org/10.1366/0003702041655421}
  {\bibfield  {journal} {\bibinfo  {journal} {Applied Spectroscopy}\ }\textbf
  {\bibinfo {volume} {58}},\ \bibinfo {pages} {986} (\bibinfo {year}
  {2004})}\BibitemShut {NoStop}%
\bibitem [{\citenamefont {Mysen}\ \emph {et~al.}(1982)\citenamefont {Mysen},
  \citenamefont {Finger}, \citenamefont {Virgo},\ and\ \citenamefont
  {A.}}]{mysen_1982}%
  \BibitemOpen
  \bibfield  {author} {\bibinfo {author} {\bibfnamefont {B.~O.}\ \bibnamefont
  {Mysen}}, \bibinfo {author} {\bibfnamefont {L.~W.}\ \bibnamefont {Finger}},
  \bibinfo {author} {\bibfnamefont {D.}~\bibnamefont {Virgo}},\ and\ \bibinfo
  {author} {\bibfnamefont {S.~F.}\ \bibnamefont {A.}},\ }\bibfield  {title}
  {\bibinfo {title} {Curve-fitting of raman spectra of silicate glasses},\
  }\href@noop {} {\bibfield  {journal} {\bibinfo  {journal} {American
  Mineralogist}\ }\textbf {\bibinfo {volume} {67}},\ \bibinfo {pages} {686}
  (\bibinfo {year} {1982})}\BibitemShut {NoStop}%
\bibitem [{\citenamefont {Gouadec}\ and\ \citenamefont
  {Colomban}(2007)}]{gouadec_raman_2007}%
  \BibitemOpen
  \bibfield  {author} {\bibinfo {author} {\bibfnamefont {G.}~\bibnamefont
  {Gouadec}}\ and\ \bibinfo {author} {\bibfnamefont {P.}~\bibnamefont
  {Colomban}},\ }\bibfield  {title} {\bibinfo {title} {Raman spectroscopy of
  nanomaterials: How spectra relate to disorder, particle size and mechanical
  properties},\ }\href {https://doi.org/10.1016/j.pcrysgrow.2007.01.001}
  {\bibfield  {journal} {\bibinfo  {journal} {Progress in Crystal Growth and
  Characterization of Materials}\ }\textbf {\bibinfo {volume} {53}},\ \bibinfo
  {pages} {1} (\bibinfo {year} {2007})}\BibitemShut {NoStop}%
\bibitem [{\citenamefont {Cuscó}\ \emph {et~al.}(2007)\citenamefont {Cuscó},
  \citenamefont {Alarcón-Lladó}, \citenamefont {Ibáñez}, \citenamefont
  {Artús}, \citenamefont {Jiménez}, \citenamefont {Wang},\ and\ \citenamefont
  {Callahan}}]{cusco_temperature_2007}%
  \BibitemOpen
  \bibfield  {author} {\bibinfo {author} {\bibfnamefont {R.}~\bibnamefont
  {Cuscó}}, \bibinfo {author} {\bibfnamefont {E.}~\bibnamefont
  {Alarcón-Lladó}}, \bibinfo {author} {\bibfnamefont {J.}~\bibnamefont
  {Ibáñez}}, \bibinfo {author} {\bibfnamefont {L.}~\bibnamefont {Artús}},
  \bibinfo {author} {\bibfnamefont {J.}~\bibnamefont {Jiménez}}, \bibinfo
  {author} {\bibfnamefont {B.}~\bibnamefont {Wang}},\ and\ \bibinfo {author}
  {\bibfnamefont {M.~J.}\ \bibnamefont {Callahan}},\ }\bibfield  {title}
  {\bibinfo {title} {Temperature dependence of raman scattering in zno},\
  }\href {https://doi.org/10.1103/PhysRevB.75.165202} {\bibfield  {journal}
  {\bibinfo  {journal} {Physical Review B}\ }\textbf {\bibinfo {volume} {75}},\
  \bibinfo {pages} {165202} (\bibinfo {year} {2007})}\BibitemShut {NoStop}%
\bibitem [{\citenamefont {Schwarz}\ and\ \citenamefont {Maier}(5
  01)}]{schwarz_asymmetric_1997}%
  \BibitemOpen
  \bibfield  {author} {\bibinfo {author} {\bibfnamefont {U.~T.}\ \bibnamefont
  {Schwarz}}\ and\ \bibinfo {author} {\bibfnamefont {M.}~\bibnamefont
  {Maier}},\ }\bibfield  {title} {\bibinfo {title} {Asymmetric raman lines
  caused by an anharmonic lattice potential in lithium niobate},\ }\href
  {https://doi.org/10.1103/PhysRevB.55.11041} {\bibfield  {journal} {\bibinfo
  {journal} {Physical Review B}\ }\textbf {\bibinfo {volume} {55}},\ \bibinfo
  {pages} {11041} (\bibinfo {year} {1997-05-01})}\BibitemShut {NoStop}%
\bibitem [{\citenamefont {Adu}\ \emph {et~al.}(2006)\citenamefont {Adu},
  \citenamefont {Xiong}, \citenamefont {Gutierrez}, \citenamefont {Chen},\ and\
  \citenamefont {Eklund}}]{adu_raman_2006}%
  \BibitemOpen
  \bibfield  {author} {\bibinfo {author} {\bibfnamefont {K.}~\bibnamefont
  {Adu}}, \bibinfo {author} {\bibfnamefont {Q.}~\bibnamefont {Xiong}}, \bibinfo
  {author} {\bibfnamefont {H.}~\bibnamefont {Gutierrez}}, \bibinfo {author}
  {\bibfnamefont {G.}~\bibnamefont {Chen}},\ and\ \bibinfo {author}
  {\bibfnamefont {P.}~\bibnamefont {Eklund}},\ }\bibfield  {title} {\bibinfo
  {title} {Raman scattering as a probe of phonon confinement and surface
  optical modes in semiconducting nanowires},\ }\href
  {https://doi.org/10.1007/s00339-006-3716-8} {\bibfield  {journal} {\bibinfo
  {journal} {Applied Physics A}\ }\textbf {\bibinfo {volume} {85}},\ \bibinfo
  {pages} {287} (\bibinfo {year} {2006})}\BibitemShut {NoStop}%
\bibitem [{\citenamefont {Nemanich}\ \emph {et~al.}(1981)\citenamefont
  {Nemanich}, \citenamefont {Solin},\ and\ \citenamefont
  {Martin}}]{nemanich_light_1981}%
  \BibitemOpen
  \bibfield  {author} {\bibinfo {author} {\bibfnamefont {R.~J.}\ \bibnamefont
  {Nemanich}}, \bibinfo {author} {\bibfnamefont {S.~A.}\ \bibnamefont
  {Solin}},\ and\ \bibinfo {author} {\bibfnamefont {R.~M.}\ \bibnamefont
  {Martin}},\ }\bibfield  {title} {\bibinfo {title} {{Light scattering study of
  boron nitride microcrystals}},\ }\href
  {https://doi.org/10.1103/PhysRevB.23.6348} {\bibfield  {journal} {\bibinfo
  {journal} {Physical Review B}\ }\textbf {\bibinfo {volume} {23}},\ \bibinfo
  {pages} {6348} (\bibinfo {year} {1981})}\BibitemShut {NoStop}%
\bibitem [{\citenamefont {Korepanov}\ and\ \citenamefont
  {Hamaguchi}(2017)}]{korepanov_quantum-chemical_nodate}%
  \BibitemOpen
  \bibfield  {author} {\bibinfo {author} {\bibfnamefont {V.~I.}\ \bibnamefont
  {Korepanov}}\ and\ \bibinfo {author} {\bibfnamefont {H.~O.}\ \bibnamefont
  {Hamaguchi}},\ }\bibfield  {title} {\bibinfo {title} {{Quantum-chemical
  perspective of nanoscale Raman spectroscopy with the three-dimensional phonon
  confinement model}},\ }\href {https://doi.org/10.1002/jrs.5132} {\bibfield
  {journal} {\bibinfo  {journal} {Journal of Raman Spectroscopy}\ }\textbf
  {\bibinfo {volume} {48}},\ \bibinfo {pages} {842} (\bibinfo {year}
  {2017})}\BibitemShut {NoStop}%
\bibitem [{\citenamefont {Martins~Ferreira}\ \emph {et~al.}(2010)\citenamefont
  {Martins~Ferreira}, \citenamefont {Moutinho}, \citenamefont {Stavale},
  \citenamefont {Lucchese}, \citenamefont {Capaz}, \citenamefont {Achete},\
  and\ \citenamefont {Jorio}}]{martins_ferreira_evolution_2010}%
  \BibitemOpen
  \bibfield  {author} {\bibinfo {author} {\bibfnamefont {E.~H.}\ \bibnamefont
  {Martins~Ferreira}}, \bibinfo {author} {\bibfnamefont {M.~V.~O.}\
  \bibnamefont {Moutinho}}, \bibinfo {author} {\bibfnamefont {F.}~\bibnamefont
  {Stavale}}, \bibinfo {author} {\bibfnamefont {M.~M.}\ \bibnamefont
  {Lucchese}}, \bibinfo {author} {\bibfnamefont {R.~B.}\ \bibnamefont {Capaz}},
  \bibinfo {author} {\bibfnamefont {C.~A.}\ \bibnamefont {Achete}},\ and\
  \bibinfo {author} {\bibfnamefont {A.}~\bibnamefont {Jorio}},\ }\bibfield
  {title} {\bibinfo {title} {Evolution of the raman spectra from single-, few-,
  and many-layer graphene with increasing disorder},\ }\href
  {https://doi.org/10.1103/PhysRevB.82.125429} {\bibfield  {journal} {\bibinfo
  {journal} {Physical Review B}\ }\textbf {\bibinfo {volume} {82}},\ \bibinfo
  {pages} {125429} (\bibinfo {year} {2010})}\BibitemShut {NoStop}%
\bibitem [{\citenamefont {Huang}\ \emph {et~al.}(2009)\citenamefont {Huang},
  \citenamefont {Yan}, \citenamefont {Chen}, \citenamefont {Song},
  \citenamefont {Heinz},\ and\ \citenamefont {Hone}}]{huang_phonon_2009}%
  \BibitemOpen
  \bibfield  {author} {\bibinfo {author} {\bibfnamefont {M.}~\bibnamefont
  {Huang}}, \bibinfo {author} {\bibfnamefont {H.}~\bibnamefont {Yan}}, \bibinfo
  {author} {\bibfnamefont {C.}~\bibnamefont {Chen}}, \bibinfo {author}
  {\bibfnamefont {D.}~\bibnamefont {Song}}, \bibinfo {author} {\bibfnamefont
  {T.~F.}\ \bibnamefont {Heinz}},\ and\ \bibinfo {author} {\bibfnamefont
  {J.}~\bibnamefont {Hone}},\ }\bibfield  {title} {\bibinfo {title} {{Phonon
  softening and crystallographic orientation of strained graphene studied by
  Raman spectroscopy}},\ }\href {https://doi.org/10.1073/pnas.0811754106}
  {\bibfield  {journal} {\bibinfo  {journal} {Proceedings of the National
  Academy of Sciences of the United States of America}\ }\textbf {\bibinfo
  {volume} {106}},\ \bibinfo {pages} {7304} (\bibinfo {year}
  {2009})}\BibitemShut {NoStop}%
\bibitem [{\citenamefont {Lehnert}\ \emph {et~al.}(1997)\citenamefont
  {Lehnert}, \citenamefont {Hendra}, \citenamefont {Everall},\ and\
  \citenamefont {Clayden}}]{lehnert_comparative_1997}%
  \BibitemOpen
  \bibfield  {author} {\bibinfo {author} {\bibfnamefont {R.}~\bibnamefont
  {Lehnert}}, \bibinfo {author} {\bibfnamefont {P.}~\bibnamefont {Hendra}},
  \bibinfo {author} {\bibfnamefont {N.}~\bibnamefont {Everall}},\ and\ \bibinfo
  {author} {\bibfnamefont {N.}~\bibnamefont {Clayden}},\ }\bibfield  {title}
  {\bibinfo {title} {Comparative quantitative study on the crystallinity of
  poly(tetrafluoroethylene) including raman, infra-red and
  \textsuperscript{19}\uppercase{F} nuclear magnetic resonance spectroscopy},\
  }\href {https://doi.org/10.1016/S0032-3861(96)00684-2} {\bibfield  {journal}
  {\bibinfo  {journal} {Polymer}\ }\textbf {\bibinfo {volume} {38}},\ \bibinfo
  {pages} {1521} (\bibinfo {year} {1997})}\BibitemShut {NoStop}%
\bibitem [{\citenamefont {Scott}(1968)}]{scott_evidence_1968}%
  \BibitemOpen
  \bibfield  {author} {\bibinfo {author} {\bibfnamefont {J.~F.}\ \bibnamefont
  {Scott}},\ }\bibfield  {title} {\bibinfo {title} {Evidence of coupling
  between one- and two-phonon excitations in quartz},\ }\href
  {https://doi.org/10.1103/PhysRevLett.21.907} {\bibfield  {journal} {\bibinfo
  {journal} {Physical Review Letters}\ }\textbf {\bibinfo {volume} {21}},\
  \bibinfo {pages} {907} (\bibinfo {year} {1968})}\BibitemShut {NoStop}%
\bibitem [{\citenamefont {Chaves}\ \emph
  {et~al.}(1974{\natexlab{a}})\citenamefont {Chaves}, \citenamefont {Katiyar},\
  and\ \citenamefont {Porto}}]{Chaves1974}%
  \BibitemOpen
  \bibfield  {author} {\bibinfo {author} {\bibfnamefont {A.}~\bibnamefont
  {Chaves}}, \bibinfo {author} {\bibfnamefont {R.~S.}\ \bibnamefont
  {Katiyar}},\ and\ \bibinfo {author} {\bibfnamefont {S.~P.}\ \bibnamefont
  {Porto}},\ }\bibfield  {title} {\bibinfo {title} {{Coupled modes with A1
  symmetry in tetragonal BaTiO3}},\ }\href
  {https://doi.org/10.1103/PhysRevB.10.3522} {\bibfield  {journal} {\bibinfo
  {journal} {Physical Review B}\ }\textbf {\bibinfo {volume} {10}},\ \bibinfo
  {pages} {3522} (\bibinfo {year} {1974}{\natexlab{a}})}\BibitemShut {NoStop}%
\bibitem [{\citenamefont {Scott}(1970)}]{scott_hybrid_1970}%
  \BibitemOpen
  \bibfield  {author} {\bibinfo {author} {\bibfnamefont {J.~F.}\ \bibnamefont
  {Scott}},\ }\bibfield  {title} {\bibinfo {title} {{Hybrid Phonons and
  Anharmonic Interactions in AlPO4}},\ }\href
  {https://doi.org/10.1103/PhysRevLett.24.1107} {\bibfield  {journal} {\bibinfo
   {journal} {Physical Review Letters}\ }\textbf {\bibinfo {volume} {24}},\
  \bibinfo {pages} {1107} (\bibinfo {year} {1970})}\BibitemShut {NoStop}%
\bibitem [{\citenamefont {Scott}(1974)}]{scott_soft-mode_1974}%
  \BibitemOpen
  \bibfield  {author} {\bibinfo {author} {\bibfnamefont {J.~F.}\ \bibnamefont
  {Scott}},\ }\bibfield  {title} {\bibinfo {title} {{Soft-mode spectroscopy:
  Experimental studies of structural phase transitions}},\ }\href
  {https://doi.org/10.1103/RevModPhys.46.83} {\bibfield  {journal} {\bibinfo
  {journal} {Reviews of Modern Physics}\ }\textbf {\bibinfo {volume} {46}},\
  \bibinfo {pages} {83} (\bibinfo {year} {1974})}\BibitemShut {NoStop}%
\bibitem [{\citenamefont {She}\ \emph {et~al.}(1972)\citenamefont {She},
  \citenamefont {Broberg}, \citenamefont {Wall},\ and\ \citenamefont
  {Edwards}}]{she_effect_1972}%
  \BibitemOpen
  \bibfield  {author} {\bibinfo {author} {\bibfnamefont {C.~Y.}\ \bibnamefont
  {She}}, \bibinfo {author} {\bibfnamefont {T.~W.}\ \bibnamefont {Broberg}},
  \bibinfo {author} {\bibfnamefont {L.~S.}\ \bibnamefont {Wall}},\ and\
  \bibinfo {author} {\bibfnamefont {D.~F.}\ \bibnamefont {Edwards}},\
  }\bibfield  {title} {\bibinfo {title} {Effect of proton-phonon coupling on
  the ferroelectric mode in $\mathrm{KH_2PO_4}$},\ }\href
  {https://doi.org/10.1103/PhysRevB.6.1847} {\bibfield  {journal} {\bibinfo
  {journal} {Physical Review B}\ }\textbf {\bibinfo {volume} {6}},\ \bibinfo
  {pages} {1847} (\bibinfo {year} {1972})}\BibitemShut {NoStop}%
\bibitem [{\citenamefont {Katiyar}\ \emph {et~al.}(1971)\citenamefont
  {Katiyar}, \citenamefont {Ryan},\ and\ \citenamefont
  {Scott}}]{katiyar_proton-phonon_1971}%
  \BibitemOpen
  \bibfield  {author} {\bibinfo {author} {\bibfnamefont {R.~S.}\ \bibnamefont
  {Katiyar}}, \bibinfo {author} {\bibfnamefont {J.~F.}\ \bibnamefont {Ryan}},\
  and\ \bibinfo {author} {\bibfnamefont {J.~F.}\ \bibnamefont {Scott}},\
  }\bibfield  {title} {\bibinfo {title} {Proton-phonon coupling in
  $\mathrm{CsH_2AsO_4}$ and $\mathrm{KH_2AsO_4}$},\ }\href
  {https://doi.org/10.1103/PhysRevB.4.2635} {\bibfield  {journal} {\bibinfo
  {journal} {Physical Review B}\ }\textbf {\bibinfo {volume} {4}},\ \bibinfo
  {pages} {2635} (\bibinfo {year} {1971})}\BibitemShut {NoStop}%
\bibitem [{\citenamefont {Rousseau}\ and\ \citenamefont
  {Porto}(1968)}]{rousseau_auger-like_1968}%
  \BibitemOpen
  \bibfield  {author} {\bibinfo {author} {\bibfnamefont {D.~L.}\ \bibnamefont
  {Rousseau}}\ and\ \bibinfo {author} {\bibfnamefont {S.~P.~S.}\ \bibnamefont
  {Porto}},\ }\bibfield  {title} {\bibinfo {title} {Auger-like resonant
  interference in raman scattering from one- and two-phonon states of
  $\mathrm{BaTiO_3}$},\ }\href {https://doi.org/10.1103/PhysRevLett.20.1354}
  {\bibfield  {journal} {\bibinfo  {journal} {Physical Review Letters}\
  }\textbf {\bibinfo {volume} {20}},\ \bibinfo {pages} {1354} (\bibinfo {year}
  {1968})}\BibitemShut {NoStop}%
\bibitem [{\citenamefont {Ager}\ \emph {et~al.}(1995)\citenamefont {Ager},
  \citenamefont {Walukiewicz}, \citenamefont {{McCluskey}}, \citenamefont
  {Plano},\ and\ \citenamefont {Landstrass}}]{ager_fano_1995}%
  \BibitemOpen
  \bibfield  {author} {\bibinfo {author} {\bibfnamefont {J.~W.}\ \bibnamefont
  {Ager}}, \bibinfo {author} {\bibfnamefont {W.}~\bibnamefont {Walukiewicz}},
  \bibinfo {author} {\bibfnamefont {M.}~\bibnamefont {{McCluskey}}}, \bibinfo
  {author} {\bibfnamefont {M.~A.}\ \bibnamefont {Plano}},\ and\ \bibinfo
  {author} {\bibfnamefont {M.~I.}\ \bibnamefont {Landstrass}},\ }\bibfield
  {title} {\bibinfo {title} {Fano interference of the raman phonon in heavily
  boron‐doped diamond films grown by chemical vapor deposition},\ }\href
  {https://doi.org/10.1063/1.114031} {\bibfield  {journal} {\bibinfo  {journal}
  {Applied Physics Letters}\ }\textbf {\bibinfo {volume} {66}},\ \bibinfo
  {pages} {616} (\bibinfo {year} {1995})}\BibitemShut {NoStop}%
\bibitem [{\citenamefont {Burke}\ \emph {et~al.}(2010)\citenamefont {Burke},
  \citenamefont {Chan}, \citenamefont {Williams}, \citenamefont {Wu},
  \citenamefont {Puretzky},\ and\ \citenamefont {Geohegan}}]{burke_raman_2010}%
  \BibitemOpen
  \bibfield  {author} {\bibinfo {author} {\bibfnamefont {B.~G.}\ \bibnamefont
  {Burke}}, \bibinfo {author} {\bibfnamefont {J.}~\bibnamefont {Chan}},
  \bibinfo {author} {\bibfnamefont {K.~A.}\ \bibnamefont {Williams}}, \bibinfo
  {author} {\bibfnamefont {Z.}~\bibnamefont {Wu}}, \bibinfo {author}
  {\bibfnamefont {A.~A.}\ \bibnamefont {Puretzky}},\ and\ \bibinfo {author}
  {\bibfnamefont {D.~B.}\ \bibnamefont {Geohegan}},\ }\bibfield  {title}
  {\bibinfo {title} {Raman study of fano interference in p-type doped
  silicon},\ }\href {https://doi.org/10.1002/jrs.2614} {\bibfield  {journal}
  {\bibinfo  {journal} {Journal of Raman Spectroscopy}\ }\textbf {\bibinfo
  {volume} {41}},\ \bibinfo {pages} {1759} (\bibinfo {year} {2010})},\ \Eprint
  {https://arxiv.org/abs/0910.5244 [cond-mat, physics:physics]} {0910.5244
  [cond-mat, physics:physics]} \BibitemShut {NoStop}%
\bibitem [{\citenamefont {Magidson}\ and\ \citenamefont
  {Beserman}(2002)}]{magidson_fano-type_2002}%
  \BibitemOpen
  \bibfield  {author} {\bibinfo {author} {\bibfnamefont {V.}~\bibnamefont
  {Magidson}}\ and\ \bibinfo {author} {\bibfnamefont {R.}~\bibnamefont
  {Beserman}},\ }\bibfield  {title} {\bibinfo {title} {Fano-type interference
  in the raman spectrum of photoexcited si},\ }\href
  {https://doi.org/10.1103/PhysRevB.66.195206} {\bibfield  {journal} {\bibinfo
  {journal} {Physical Review B}\ }\textbf {\bibinfo {volume} {66}},\ \bibinfo
  {pages} {195206} (\bibinfo {year} {2002})}\BibitemShut {NoStop}%
\bibitem [{\citenamefont {Fano}(1961)}]{fano_effects_nodate}%
  \BibitemOpen
  \bibfield  {author} {\bibinfo {author} {\bibfnamefont {U.}~\bibnamefont
  {Fano}},\ }\bibfield  {title} {\bibinfo {title} {{Effects of configuration
  interaction on intensities and phase shifts}},\ }\href
  {https://doi.org/10.1103/PhysRev.124.1866} {\bibfield  {journal} {\bibinfo
  {journal} {Physical Review}\ }\textbf {\bibinfo {volume} {124}},\ \bibinfo
  {pages} {1866} (\bibinfo {year} {1961})}\BibitemShut {NoStop}%
\bibitem [{\citenamefont {Breit}\ and\ \citenamefont
  {Wigner}(1936)}]{breit_capture_nodate}%
  \BibitemOpen
  \bibfield  {author} {\bibinfo {author} {\bibfnamefont {G.}~\bibnamefont
  {Breit}}\ and\ \bibinfo {author} {\bibfnamefont {E.}~\bibnamefont {Wigner}},\
  }\bibfield  {title} {\bibinfo {title} {{Capture of slow neutrons}},\ }\href
  {https://doi.org/10.1103/PhysRev.49.519} {\bibfield  {journal} {\bibinfo
  {journal} {Physical Review}\ }\textbf {\bibinfo {volume} {49}},\ \bibinfo
  {pages} {519} (\bibinfo {year} {1936})}\BibitemShut {NoStop}%
\bibitem [{\citenamefont {Brown}\ \emph {et~al.}(2001)\citenamefont {Brown},
  \citenamefont {Jorio}, \citenamefont {Corio}, \citenamefont {Dresselhaus},
  \citenamefont {Dresselhaus}, \citenamefont {Saito},\ and\ \citenamefont
  {Kneipp}}]{brown_origin_2001}%
  \BibitemOpen
  \bibfield  {author} {\bibinfo {author} {\bibfnamefont {S.~D.~M.}\
  \bibnamefont {Brown}}, \bibinfo {author} {\bibfnamefont {A.}~\bibnamefont
  {Jorio}}, \bibinfo {author} {\bibfnamefont {P.}~\bibnamefont {Corio}},
  \bibinfo {author} {\bibfnamefont {M.~S.}\ \bibnamefont {Dresselhaus}},
  \bibinfo {author} {\bibfnamefont {G.}~\bibnamefont {Dresselhaus}}, \bibinfo
  {author} {\bibfnamefont {R.}~\bibnamefont {Saito}},\ and\ \bibinfo {author}
  {\bibfnamefont {K.}~\bibnamefont {Kneipp}},\ }\bibfield  {title} {\bibinfo
  {title} {Origin of the breit-wigner-fano lineshape of the tangential
  \textit{G} -band feature of metallic carbon nanotubes},\ }\href
  {https://doi.org/10.1103/PhysRevB.63.155414} {\bibfield  {journal} {\bibinfo
  {journal} {Physical Review B}\ }\textbf {\bibinfo {volume} {63}},\ \bibinfo
  {pages} {155414} (\bibinfo {year} {2001})}\BibitemShut {NoStop}%
\bibitem [{\citenamefont {Gordeev}\ \emph {et~al.}(2016)\citenamefont
  {Gordeev}, \citenamefont {Setaro}, \citenamefont {Glaeske}, \citenamefont
  {J{\"{u}}rgensen},\ and\ \citenamefont {Reich}}]{Gordeev2016a}%
  \BibitemOpen
  \bibfield  {author} {\bibinfo {author} {\bibfnamefont {G.}~\bibnamefont
  {Gordeev}}, \bibinfo {author} {\bibfnamefont {A.}~\bibnamefont {Setaro}},
  \bibinfo {author} {\bibfnamefont {M.}~\bibnamefont {Glaeske}}, \bibinfo
  {author} {\bibfnamefont {S.}~\bibnamefont {J{\"{u}}rgensen}},\ and\ \bibinfo
  {author} {\bibfnamefont {S.}~\bibnamefont {Reich}},\ }\bibfield  {title}
  {\bibinfo {title} {{Doping in covalently functionalized carbon nanotubes: A
  Raman scattering study}},\ }\href {https://doi.org/10.1002/pssb.201600636}
  {\bibfield  {journal} {\bibinfo  {journal} {Physica Status Solidi (B)}\
  }\textbf {\bibinfo {volume} {253}},\ \bibinfo {pages} {2461} (\bibinfo {year}
  {2016})}\BibitemShut {NoStop}%
\bibitem [{\citenamefont {Barker}\ and\ \citenamefont
  {Hopfield}(1964)}]{barker_coupled-optical-phonon-mode_1964}%
  \BibitemOpen
  \bibfield  {author} {\bibinfo {author} {\bibfnamefont {A.~S.}\ \bibnamefont
  {Barker}}\ and\ \bibinfo {author} {\bibfnamefont {J.~J.}\ \bibnamefont
  {Hopfield}},\ }\bibfield  {title} {\bibinfo {title}
  {Coupled-optical-phonon-mode theory of the infrared dispersion in
  $\mathrm{BaTiO_3}$ , $\mathrm{SrTiO_3}$ , and $\mathrm{KTaO_3}$},\ }\href
  {https://doi.org/10.1103/PhysRev.135.A1732} {\bibfield  {journal} {\bibinfo
  {journal} {Physical Review}\ }\textbf {\bibinfo {volume} {135}},\ \bibinfo
  {pages} {A1732} (\bibinfo {year} {1964})}\BibitemShut {NoStop}%
\bibitem [{\citenamefont {Zawadowski}\ and\ \citenamefont
  {Ruvalds}(1970)}]{ruvalds_indirect_1970}%
  \BibitemOpen
  \bibfield  {author} {\bibinfo {author} {\bibfnamefont {A.}~\bibnamefont
  {Zawadowski}}\ and\ \bibinfo {author} {\bibfnamefont {J.}~\bibnamefont
  {Ruvalds}},\ }\bibfield  {title} {\bibinfo {title} {{Indirect Coupling and
  Antiresonance of Two Optic Phonons}},\ }\href
  {https://doi.org/10.1103/PhysRevLett.24.1111} {\bibfield  {journal} {\bibinfo
   {journal} {Physical Review Letters}\ }\textbf {\bibinfo {volume} {24}},\
  \bibinfo {pages} {1111} (\bibinfo {year} {1970})}\BibitemShut {NoStop}%
\bibitem [{\citenamefont {Scalabrin}\ \emph {et~al.}(1977)\citenamefont
  {Scalabrin}, \citenamefont {Chaves}, \citenamefont {Shim},\ and\
  \citenamefont {Porto}}]{scalabrin_temperature_1977}%
  \BibitemOpen
  \bibfield  {author} {\bibinfo {author} {\bibfnamefont {A.}~\bibnamefont
  {Scalabrin}}, \bibinfo {author} {\bibfnamefont {A.~S.}\ \bibnamefont
  {Chaves}}, \bibinfo {author} {\bibfnamefont {D.~S.}\ \bibnamefont {Shim}},\
  and\ \bibinfo {author} {\bibfnamefont {S.~P.}\ \bibnamefont {Porto}},\
  }\bibfield  {title} {\bibinfo {title} {{Temperature dependence of the A1 and
  E optical phonons in BaTiO3}},\ }\href
  {https://doi.org/10.1002/pssb.2220790240} {\bibfield  {journal} {\bibinfo
  {journal} {Physica Status Solidi (B)}\ }\textbf {\bibinfo {volume} {79}},\
  \bibinfo {pages} {731} (\bibinfo {year} {1977})}\BibitemShut {NoStop}%
\bibitem [{\citenamefont {Lagakos}\ and\ \citenamefont
  {Cummins}(1974)}]{lagakos_preliminary_1974}%
  \BibitemOpen
  \bibfield  {author} {\bibinfo {author} {\bibfnamefont {N.}~\bibnamefont
  {Lagakos}}\ and\ \bibinfo {author} {\bibfnamefont {H.~Z.}\ \bibnamefont
  {Cummins}},\ }\bibfield  {title} {\bibinfo {title} {Preliminary observation
  of a central peak in the light-scattering spectrum of $kh_2po_4$},\ }\href
  {https://doi.org/10.1103/PhysRevB.10.1063} {\bibfield  {journal} {\bibinfo
  {journal} {Physical Review B}\ }\textbf {\bibinfo {volume} {10}},\ \bibinfo
  {pages} {1063} (\bibinfo {year} {1974})}\BibitemShut {NoStop}%
\bibitem [{\citenamefont {Bozinis}\ and\ \citenamefont
  {Hurrell}(1976)}]{bozinis_optical_1976}%
  \BibitemOpen
  \bibfield  {author} {\bibinfo {author} {\bibfnamefont {D.~G.}\ \bibnamefont
  {Bozinis}}\ and\ \bibinfo {author} {\bibfnamefont {J.~P.}\ \bibnamefont
  {Hurrell}},\ }\bibfield  {title} {\bibinfo {title} {Optical modes and
  dielectric properties of ferroelectric orthorhombic $\mathrm{KNbO_3}$},\
  }\href {https://doi.org/10.1103/PhysRevB.13.3109} {\bibfield  {journal}
  {\bibinfo  {journal} {Physical Review B}\ }\textbf {\bibinfo {volume} {13}},\
  \bibinfo {pages} {3109} (\bibinfo {year} {1976})}\BibitemShut {NoStop}%
\bibitem [{\citenamefont {Benyuan}\ \emph {et~al.}(1981)\citenamefont
  {Benyuan}, \citenamefont {Copic},\ and\ \citenamefont
  {Cummins}}]{benyuan_soft_1981}%
  \BibitemOpen
  \bibfield  {author} {\bibinfo {author} {\bibfnamefont {G.}~\bibnamefont
  {Benyuan}}, \bibinfo {author} {\bibfnamefont {M.}~\bibnamefont {Copic}},\
  and\ \bibinfo {author} {\bibfnamefont {H.~Z.}\ \bibnamefont {Cummins}},\
  }\bibfield  {title} {\bibinfo {title} {Soft acoustic mode in ferroelastic
  $\mathrm{BiVO_4}$},\ }\href {https://doi.org/10.1103/PhysRevB.24.4098}
  {\bibfield  {journal} {\bibinfo  {journal} {Physical Review B}\ }\textbf
  {\bibinfo {volume} {24}},\ \bibinfo {pages} {4098} (\bibinfo {year}
  {1981})}\BibitemShut {NoStop}%
\bibitem [{\citenamefont {Avakyants}\ \emph {et~al.}(2004)\citenamefont
  {Avakyants}, \citenamefont {Chervyakov}, \citenamefont {Gorelik},\ and\
  \citenamefont {Sverbil'}}]{avakyants_inelastic_2004}%
  \BibitemOpen
  \bibfield  {author} {\bibinfo {author} {\bibfnamefont {L.~P.}\ \bibnamefont
  {Avakyants}}, \bibinfo {author} {\bibfnamefont {A.~V.}\ \bibnamefont
  {Chervyakov}}, \bibinfo {author} {\bibfnamefont {V.~S.}\ \bibnamefont
  {Gorelik}},\ and\ \bibinfo {author} {\bibfnamefont {P.~P.}\ \bibnamefont
  {Sverbil'}},\ }\bibfield  {title} {\bibinfo {title} {Inelastic light
  scattering near the ferroelectric phase-transition point in bismuth vanadate
  crystals},\ }\href {https://doi.org/10.1023/B:JORR.0000049086.09037.b3}
  {\bibfield  {journal} {\bibinfo  {journal} {Journal of Russian Laser
  Research}\ }\textbf {\bibinfo {volume} {25}},\ \bibinfo {pages} {535}
  (\bibinfo {year} {2004})}\BibitemShut {NoStop}%
\bibitem [{\citenamefont {Pellicer-Porres}\ \emph {et~al.}(2018)\citenamefont
  {Pellicer-Porres}, \citenamefont {Vázquez-Socorro}, \citenamefont
  {López-Moreno}, \citenamefont {Muñoz}, \citenamefont
  {Rodríguez-Hernández}, \citenamefont {Martínez-García}, \citenamefont
  {Achary}, \citenamefont {Rettie},\ and\ \citenamefont
  {Mullins}}]{pellicer-porres_phase_2018}%
  \BibitemOpen
  \bibfield  {author} {\bibinfo {author} {\bibfnamefont {J.}~\bibnamefont
  {Pellicer-Porres}}, \bibinfo {author} {\bibfnamefont {D.}~\bibnamefont
  {Vázquez-Socorro}}, \bibinfo {author} {\bibfnamefont {S.}~\bibnamefont
  {López-Moreno}}, \bibinfo {author} {\bibfnamefont {A.}~\bibnamefont
  {Muñoz}}, \bibinfo {author} {\bibfnamefont {P.}~\bibnamefont
  {Rodríguez-Hernández}}, \bibinfo {author} {\bibfnamefont {D.}~\bibnamefont
  {Martínez-García}}, \bibinfo {author} {\bibfnamefont {S.~N.}\ \bibnamefont
  {Achary}}, \bibinfo {author} {\bibfnamefont {A.~J.~E.}\ \bibnamefont
  {Rettie}},\ and\ \bibinfo {author} {\bibfnamefont {C.~B.}\ \bibnamefont
  {Mullins}},\ }\bibfield  {title} {\bibinfo {title} {Phase transition
  systematics in $\mathrm{BiVO_4}$ by means of high-pressure–high-temperature
  raman experiments},\ }\href {https://doi.org/10.1103/PhysRevB.98.214109}
  {\bibfield  {journal} {\bibinfo  {journal} {Physical Review B}\ }\textbf
  {\bibinfo {volume} {98}},\ \bibinfo {pages} {214109} (\bibinfo {year}
  {2018})}\BibitemShut {NoStop}%
\bibitem [{\citenamefont {Takagi}(1983)}]{takagi_coupled_1983}%
  \BibitemOpen
  \bibfield  {author} {\bibinfo {author} {\bibfnamefont {Y.}~\bibnamefont
  {Takagi}},\ }\bibfield  {title} {\bibinfo {title} {Coupled modes analysis of
  raman spectra: Application to the case of a $\mathrm{KH2PO4}$ crystal in the
  paraelectric phase},\ }\href {https://doi.org/10.1080/00150198308225271}
  {\bibfield  {journal} {\bibinfo  {journal} {Ferroelectrics}\ }\textbf
  {\bibinfo {volume} {46}},\ \bibinfo {pages} {245} (\bibinfo {year}
  {1983})}\BibitemShut {NoStop}%
\bibitem [{\citenamefont {Lowndes}\ \emph {et~al.}(1974)\citenamefont
  {Lowndes}, \citenamefont {Tornberg},\ and\ \citenamefont
  {Leung}}]{lowndes_ferroelectric_1974}%
  \BibitemOpen
  \bibfield  {author} {\bibinfo {author} {\bibfnamefont {R.~P.}\ \bibnamefont
  {Lowndes}}, \bibinfo {author} {\bibfnamefont {N.~E.}\ \bibnamefont
  {Tornberg}},\ and\ \bibinfo {author} {\bibfnamefont {R.~C.}\ \bibnamefont
  {Leung}},\ }\bibfield  {title} {\bibinfo {title} {Ferroelectric mode and
  molecular structure in the hydrogen-bonded ferroelectric arsenates and their
  deuterated isomorphs},\ }\href {https://doi.org/10.1103/PhysRevB.10.911}
  {\bibfield  {journal} {\bibinfo  {journal} {Physical Review B}\ }\textbf
  {\bibinfo {volume} {10}},\ \bibinfo {pages} {911} (\bibinfo {year}
  {1974})}\BibitemShut {NoStop}%
\bibitem [{\citenamefont {Mazzacurati}\ \emph {et~al.}(1976)\citenamefont
  {Mazzacurati}, \citenamefont {Signorelli},\ and\ \citenamefont
  {Sampoli}}]{mazzacurati_indirect_1976}%
  \BibitemOpen
  \bibfield  {author} {\bibinfo {author} {\bibfnamefont {V.}~\bibnamefont
  {Mazzacurati}}, \bibinfo {author} {\bibfnamefont {G.}~\bibnamefont
  {Signorelli}},\ and\ \bibinfo {author} {\bibfnamefont {M.}~\bibnamefont
  {Sampoli}},\ }\bibfield  {title} {\bibinfo {title} {Indirect coupling of the
  soft optic phonon in $\mathrm{KH_2PO_4}$},\ }\href
  {https://doi.org/10.1002/jrs.1250040402} {\bibfield  {journal} {\bibinfo
  {journal} {Journal of Raman Spectroscopy}\ }\textbf {\bibinfo {volume} {4}},\
  \bibinfo {pages} {339} (\bibinfo {year} {1976})}\BibitemShut {NoStop}%
\bibitem [{\citenamefont {Scarparo}\ \emph {et~al.}(1978)\citenamefont
  {Scarparo}, \citenamefont {Katiyar}, \citenamefont {Srivastava},\ and\
  \citenamefont {Porto}}]{scarparo_evidence_1978}%
  \BibitemOpen
  \bibfield  {author} {\bibinfo {author} {\bibfnamefont {M.~A.~F.}\
  \bibnamefont {Scarparo}}, \bibinfo {author} {\bibfnamefont {R.~S.}\
  \bibnamefont {Katiyar}}, \bibinfo {author} {\bibfnamefont {R.}~\bibnamefont
  {Srivastava}},\ and\ \bibinfo {author} {\bibfnamefont {S.~P.~S.}\
  \bibnamefont {Porto}},\ }\bibfield  {title} {\bibinfo {title} {Evidence of
  non-linear behaviour of soft modes in crystals with kdp structure},\ }\href
  {https://doi.org/10.1002/pssb.2220900213} {\bibfield  {journal} {\bibinfo
  {journal} {Physica Status Solidi (b)}\ }\textbf {\bibinfo {volume} {90}},\
  \bibinfo {pages} {543} (\bibinfo {year} {1978})}\BibitemShut {NoStop}%
\bibitem [{\citenamefont {Kaminow}\ and\ \citenamefont
  {Damen}(1968)}]{kaminow_temperature_1968}%
  \BibitemOpen
  \bibfield  {author} {\bibinfo {author} {\bibfnamefont {I.~P.}\ \bibnamefont
  {Kaminow}}\ and\ \bibinfo {author} {\bibfnamefont {T.~C.}\ \bibnamefont
  {Damen}},\ }\bibfield  {title} {\bibinfo {title} {{Temperature dependence of
  the ferroelectric mode in KH2PO4}},\ }\href
  {https://doi.org/10.1103/PhysRevLett.20.1105} {\bibfield  {journal} {\bibinfo
   {journal} {Physical Review Letters}\ }\textbf {\bibinfo {volume} {20}},\
  \bibinfo {pages} {1105} (\bibinfo {year} {1968})}\BibitemShut {NoStop}%
\bibitem [{\citenamefont {Burns}\ and\ \citenamefont
  {Scott}(1973)}]{burns_lattice_nodate}%
  \BibitemOpen
  \bibfield  {author} {\bibinfo {author} {\bibfnamefont {G.}~\bibnamefont
  {Burns}}\ and\ \bibinfo {author} {\bibfnamefont {B.~A.}\ \bibnamefont
  {Scott}},\ }\bibfield  {title} {\bibinfo {title} {{Lattice modes in
  ferroelectric perovskites: PbTiO3}},\ }\href
  {https://doi.org/10.1103/PhysRevB.7.3088} {\bibfield  {journal} {\bibinfo
  {journal} {Physical Review B}\ }\textbf {\bibinfo {volume} {7}},\ \bibinfo
  {pages} {3088} (\bibinfo {year} {1973})}\BibitemShut {NoStop}%
\bibitem [{\citenamefont {Chaves}\ \emph
  {et~al.}(1974{\natexlab{b}})\citenamefont {Chaves}, \citenamefont {Katiyar},\
  and\ \citenamefont
  {Porto}}]{chaves_coupled_modes_with_a1_symmetry_in_tetragpdf_1974}%
  \BibitemOpen
  \bibfield  {author} {\bibinfo {author} {\bibfnamefont {A.}~\bibnamefont
  {Chaves}}, \bibinfo {author} {\bibfnamefont {R.~S.}\ \bibnamefont
  {Katiyar}},\ and\ \bibinfo {author} {\bibfnamefont {S.~P.~S.}\ \bibnamefont
  {Porto}},\ }\bibfield  {title} {\bibinfo {title} {Coupled modes with a1
  symmetry in tetragonal $\mathrm{BaTiO_3}$},\ }\href@noop {} {\bibfield
  {journal} {\bibinfo  {journal} {Physical Review B}\ }\textbf {\bibinfo
  {volume} {10}},\ \bibinfo {pages} {3522} (\bibinfo {year}
  {1974}{\natexlab{b}})}\BibitemShut {NoStop}%
\bibitem [{\citenamefont {Liu}\ \emph {et~al.}(2000)\citenamefont {Liu},
  \citenamefont {Bursill}, \citenamefont {Prawer},\ and\ \citenamefont
  {Beserman}}]{liu_temperature_2000}%
  \BibitemOpen
  \bibfield  {author} {\bibinfo {author} {\bibfnamefont {M.~S.}\ \bibnamefont
  {Liu}}, \bibinfo {author} {\bibfnamefont {L.~A.}\ \bibnamefont {Bursill}},
  \bibinfo {author} {\bibfnamefont {S.}~\bibnamefont {Prawer}},\ and\ \bibinfo
  {author} {\bibfnamefont {R.}~\bibnamefont {Beserman}},\ }\bibfield  {title}
  {\bibinfo {title} {Temperature dependence of the first-order raman phonon
  line of diamond},\ }\href {https://doi.org/10.1103/PhysRevB.61.3391}
  {\bibfield  {journal} {\bibinfo  {journal} {Physical Review B}\ }\textbf
  {\bibinfo {volume} {61}},\ \bibinfo {pages} {3391} (\bibinfo {year}
  {2000})}\BibitemShut {NoStop}%
\bibitem [{\citenamefont {Klemens}(1966)}]{klemens_anharmonic_1966}%
  \BibitemOpen
  \bibfield  {author} {\bibinfo {author} {\bibfnamefont {P.~G.}\ \bibnamefont
  {Klemens}},\ }\bibfield  {title} {\bibinfo {title} {Anharmonic decay of
  optical phonons},\ }\href {https://doi.org/10.1103/PhysRev.148.845}
  {\bibfield  {journal} {\bibinfo  {journal} {Physical Review}\ }\textbf
  {\bibinfo {volume} {148}},\ \bibinfo {pages} {845} (\bibinfo {year}
  {1966})}\BibitemShut {NoStop}%
\bibitem [{\citenamefont {Zhu}\ \emph {et~al.}(2005)\citenamefont {Zhu},
  \citenamefont {Zhang}, \citenamefont {Chen},\ and\ \citenamefont
  {Yin}}]{zhu_size_2005}%
  \BibitemOpen
  \bibfield  {author} {\bibinfo {author} {\bibfnamefont {K.-R.}\ \bibnamefont
  {Zhu}}, \bibinfo {author} {\bibfnamefont {M.-S.}\ \bibnamefont {Zhang}},
  \bibinfo {author} {\bibfnamefont {Q.}~\bibnamefont {Chen}},\ and\ \bibinfo
  {author} {\bibfnamefont {Z.}~\bibnamefont {Yin}},\ }\bibfield  {title}
  {\bibinfo {title} {Size and phonon-confinement effects on low-frequency raman
  mode of anatase $\mathrm{TiO_2}$ nanocrystal},\ }\href
  {https://doi.org/10.1016/j.physleta.2005.04.008} {\bibfield  {journal}
  {\bibinfo  {journal} {Physics Letters A}\ }\textbf {\bibinfo {volume}
  {340}},\ \bibinfo {pages} {220} (\bibinfo {year} {2005})}\BibitemShut
  {NoStop}%
\bibitem [{\citenamefont {Spanier}\ \emph {et~al.}(2001)\citenamefont
  {Spanier}, \citenamefont {Robinson}, \citenamefont {Zhang}, \citenamefont
  {Chan},\ and\ \citenamefont {Herman}}]{spanier_size-dependent_2001}%
  \BibitemOpen
  \bibfield  {author} {\bibinfo {author} {\bibfnamefont {J.~E.}\ \bibnamefont
  {Spanier}}, \bibinfo {author} {\bibfnamefont {R.~D.}\ \bibnamefont
  {Robinson}}, \bibinfo {author} {\bibfnamefont {F.}~\bibnamefont {Zhang}},
  \bibinfo {author} {\bibfnamefont {S.-W.}\ \bibnamefont {Chan}},\ and\
  \bibinfo {author} {\bibfnamefont {I.~P.}\ \bibnamefont {Herman}},\ }\bibfield
   {title} {\bibinfo {title} {Size-dependent properties of $\mathrm{CeO_{2-y}}$
  nanoparticles as studied by raman scattering},\ }\href
  {https://doi.org/10.1103/PhysRevB.64.245407} {\bibfield  {journal} {\bibinfo
  {journal} {Physical Review B}\ }\textbf {\bibinfo {volume} {64}},\ \bibinfo
  {pages} {245407} (\bibinfo {year} {2001})}\BibitemShut {NoStop}%
\bibitem [{\citenamefont {Wang}\ \emph {et~al.}(2007)\citenamefont {Wang},
  \citenamefont {Chen},\ and\ \citenamefont {Zhao}}]{wang_lattice_2007}%
  \BibitemOpen
  \bibfield  {author} {\bibinfo {author} {\bibfnamefont {D.}~\bibnamefont
  {Wang}}, \bibinfo {author} {\bibfnamefont {B.}~\bibnamefont {Chen}},\ and\
  \bibinfo {author} {\bibfnamefont {J.}~\bibnamefont {Zhao}},\ }\bibfield
  {title} {\bibinfo {title} {Lattice vibration fundamentals of nanocrystalline
  anatase: Temperature-dependent study using micro-raman scattering
  spectroscopy},\ }\href {https://doi.org/10.1063/1.2736309} {\bibfield
  {journal} {\bibinfo  {journal} {Journal of Applied Physics}\ }\textbf
  {\bibinfo {volume} {101}},\ \bibinfo {pages} {113501} (\bibinfo {year}
  {2007})}\BibitemShut {NoStop}%
\bibitem [{\citenamefont {Menéndez}\ and\ \citenamefont
  {Cardona}(1984)}]{menendez_temperature_1984}%
  \BibitemOpen
  \bibfield  {author} {\bibinfo {author} {\bibfnamefont {J.}~\bibnamefont
  {Menéndez}}\ and\ \bibinfo {author} {\bibfnamefont {M.}~\bibnamefont
  {Cardona}},\ }\bibfield  {title} {\bibinfo {title} {Temperature dependence of
  the first-order raman scattering by phonons in si, ge, and $\alpha$-sn :
  Anharmonic effects},\ }\href {https://doi.org/10.1103/PhysRevB.29.2051}
  {\bibfield  {journal} {\bibinfo  {journal} {Physical Review B}\ }\textbf
  {\bibinfo {volume} {29}},\ \bibinfo {pages} {2051} (\bibinfo {year}
  {1984})}\BibitemShut {NoStop}%
\bibitem [{\citenamefont {Meier}(2005{\natexlab{b}})}]{meier_art_2005}%
  \BibitemOpen
  \bibfield  {author} {\bibinfo {author} {\bibfnamefont {R.~J.}\ \bibnamefont
  {Meier}},\ }\bibfield  {title} {\bibinfo {title} {On art and science in
  curve-fitting vibrational spectra},\ }\href
  {https://doi.org/10.1016/j.vibspec.2005.03.003} {\bibfield  {journal}
  {\bibinfo  {journal} {Vibrational Spectroscopy}\ }\textbf {\bibinfo {volume}
  {39}},\ \bibinfo {pages} {266} (\bibinfo {year}
  {2005}{\natexlab{b}})}\BibitemShut {NoStop}%
\bibitem [{\citenamefont {Zhou}\ \emph {et~al.}(1992)\citenamefont {Zhou},
  \citenamefont {Wang}, \citenamefont {Rao}, \citenamefont {Eklund},
  \citenamefont {Dresselhaus},\ and\ \citenamefont
  {Dresselhaus}}]{zhou_raman-scattering_1992}%
  \BibitemOpen
  \bibfield  {author} {\bibinfo {author} {\bibfnamefont {P.}~\bibnamefont
  {Zhou}}, \bibinfo {author} {\bibfnamefont {K.-A.}\ \bibnamefont {Wang}},
  \bibinfo {author} {\bibfnamefont {A.~M.}\ \bibnamefont {Rao}}, \bibinfo
  {author} {\bibfnamefont {P.~C.}\ \bibnamefont {Eklund}}, \bibinfo {author}
  {\bibfnamefont {G.}~\bibnamefont {Dresselhaus}},\ and\ \bibinfo {author}
  {\bibfnamefont {M.~S.}\ \bibnamefont {Dresselhaus}},\ }\bibfield  {title}
  {\bibinfo {title} {Raman-scattering studies of homogeneous $\mathrm{K_3C_60}$
  films},\ }\href {https://doi.org/10.1103/PhysRevB.45.10838} {\bibfield
  {journal} {\bibinfo  {journal} {Physical Review B}\ }\textbf {\bibinfo
  {volume} {45}},\ \bibinfo {pages} {10838} (\bibinfo {year}
  {1992})}\BibitemShut {NoStop}%
\end{thebibliography}%


\begin{thebibliography}{0}%
\makeatletter
\providecommand \@ifxundefined [1]{%
 \@ifx{#1\undefined}
}%
\providecommand \@ifnum [1]{%
 \ifnum #1\expandafter \@firstoftwo
 \else \expandafter \@secondoftwo
 \fi
}%
\providecommand \@ifx [1]{%
 \ifx #1\expandafter \@firstoftwo
 \else \expandafter \@secondoftwo
 \fi
}%
\providecommand \natexlab [1]{#1}%
\providecommand \enquote  [1]{``#1''}%
\providecommand \bibnamefont  [1]{#1}%
\providecommand \bibfnamefont [1]{#1}%
\providecommand \citenamefont [1]{#1}%
\providecommand \href@noop [0]{\@secondoftwo}%
\providecommand \href [0]{\begingroup \@sanitize@url \@href}%
\providecommand \@href[1]{\@@startlink{#1}\@@href}%
\providecommand \@@href[1]{\endgroup#1\@@endlink}%
\providecommand \@sanitize@url [0]{\catcode `\\12\catcode `\$12\catcode
  `\&12\catcode `\#12\catcode `\^12\catcode `\_12\catcode `\%12\relax}%
\providecommand \@@startlink[1]{}%
\providecommand \@@endlink[0]{}%
\providecommand \url  [0]{\begingroup\@sanitize@url \@url }%
\providecommand \@url [1]{\endgroup\@href {#1}{\urlprefix }}%
\providecommand \urlprefix  [0]{URL }%
\providecommand \Eprint [0]{\href }%
\providecommand \doibase [0]{https://doi.org/}%
\providecommand \selectlanguage [0]{\@gobble}%
\providecommand \bibinfo  [0]{\@secondoftwo}%
\providecommand \bibfield  [0]{\@secondoftwo}%
\providecommand \translation [1]{[#1]}%
\providecommand \BibitemOpen [0]{}%
\providecommand \bibitemStop [0]{}%
\providecommand \bibitemNoStop [0]{.\EOS\space}%
\providecommand \EOS [0]{\spacefactor3000\relax}%
\providecommand \BibitemShut  [1]{\csname bibitem#1\endcsname}%
\let\auto@bib@innerbib\@empty
\end{thebibliography}%

\end{document}


\title{Modeling a polarization-dependent phonon-phonon coupling in the Raman spectrum of BiVO$_4$ over a large temperature range \\ -- \\ Supplementary information}
\author{Christina Hill}
\affiliation{Materials Research and Technology Department, Luxembourg Institute of Science and Technology, 41 rue du Brill, L-4422 Belvaux, Luxembourg}
\affiliation{Department of Physics and Materials Science, University of Luxembourg, 41 rue du Brill, L-4422 Belvaux, Luxembourg}
\author{Georgy Gordeev}
\affiliation{Department of Physics and Materials Science, University of Luxembourg, 41 rue du Brill, L-4422 Belvaux, Luxembourg}
\author{Mael Guennou}
\affiliation{Department of Physics and Materials Science, University of Luxembourg, 41 rue du Brill, L-4422 Belvaux, Luxembourg}

\maketitle

\tableofcontents
\clearpage

\section*{Coupled damped harmonic oscillator and its coupling coefficients}
\begin{figure}[h!]
\centering
 \includegraphics[width=\linewidth]{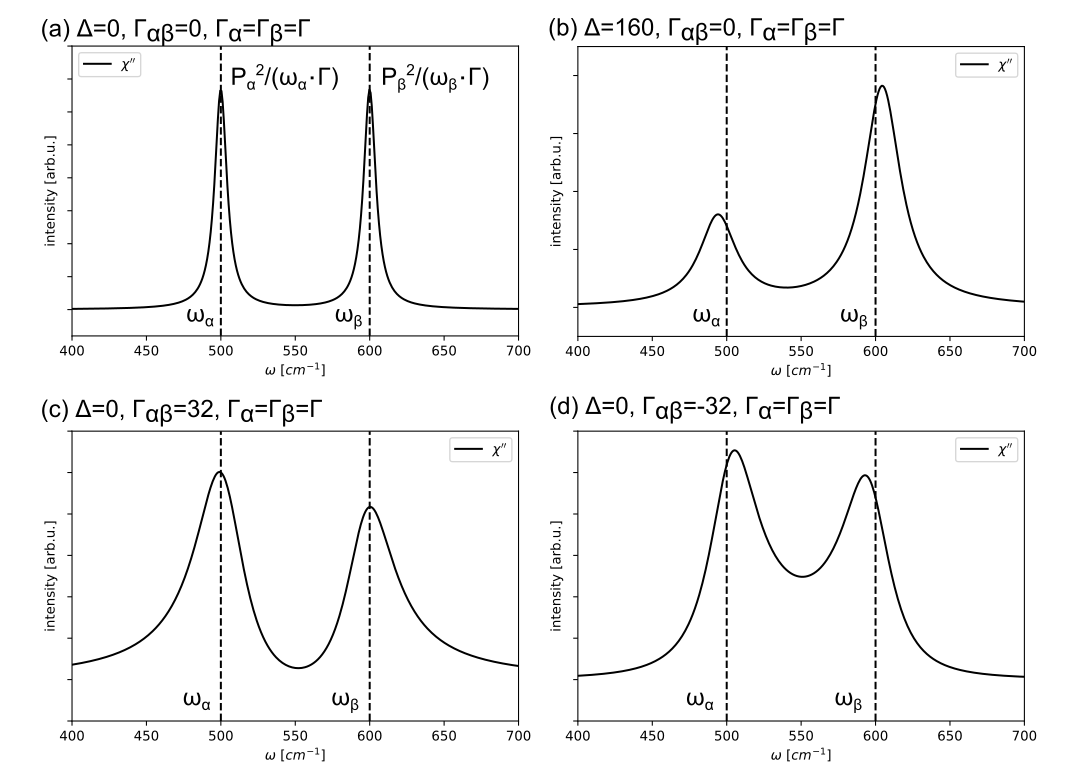}
\caption{Imaginary part of the susceptibility $\chi''(\omega)$ calculated from Greens Formalism (a) without coupling assuming $\Delta=0$ and damping $\Gamma_{\alpha\beta}=0$ (b) $\Delta$ is non-zero (c) $\Gamma_{\alpha\beta}$ is positive (d) $\Gamma_{\alpha\beta}$ is negative}
\label{fig:GreensForm}
\end{figure}

Fig.~\ref{fig:GreensForm} shows the imaginary part of the susceptibility $\chi''(\omega)$ calculated from the Greens Formalism (a) assuming no coupling, (b) $\Delta$ is non-zero, (c) $\Gamma_{\alpha\beta}$ is positive and (d) $\Gamma_{\alpha\beta}$ is negative. In case of $\Delta$ and $\Gamma_{\alpha\beta}$ being zero, the maximum intensity appears at the eigen-frequencies $\omega_{\alpha}$ and $\omega_{\beta}$ of the individual phonons and the line-shape of both phonon bands is symmetric, see Fig.~\ref{fig:GreensForm}~(a).  For non-zero $\Delta$, the maxima no longer appear at the eigen-frequencies (repulsion of modes) and there is an intensity transfer from the phonon mode $\alpha$ to $\beta$. 
Figs.~\ref{fig:GreensForm}~(c) and (d) show the imaginary part of the susceptibility assuming positive and negative $\Gamma_{\alpha\beta}$, respectively. For positive $\Gamma_{\alpha\beta}$ the two phonon modes are pushed apart. For negative $\Gamma_{\alpha\beta}$ the two phonon modes are attracted. For both pure real and pure imaginary coupling the line-shape of the phonon bands becomes asymmetric.

\clearpage

\section*{Raman modes of $A_g$ symmetry of monoclinic bismuth vanadate}

\begin{figure}[h!]
\centering
 \includegraphics[width=0.8\linewidth]{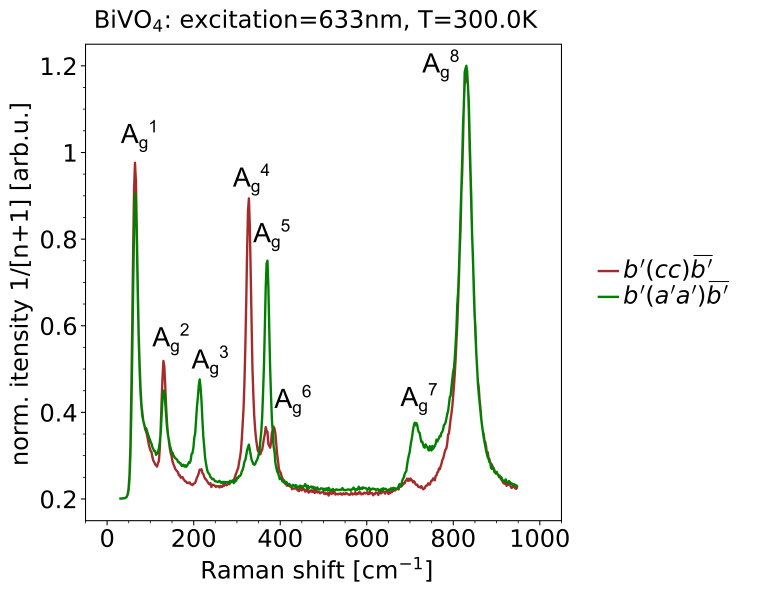}
\caption{Raman spectra of the single crystal BiVO$_4$ measured at T=\SI{300}{\K} with a laser excitation of \SI{633}{\nm} for the light polarizations $b'(cc)\bar{b}'$ and $b'(a'a')\bar{b}'$. All spectra are normalized to the $A_g^8$ mode at \SI{830}{\wn}.}
\label{fig:BVO_spec}
\end{figure}

Fig.\ref{fig:BVO_spec} shows the Raman spectrum of BiVO$_4$ measured at T=\SI{300}{\K} with a laser excitation of \SI{633}{\nm} for the light polarizations $b'(cc)\bar{b}'$ and $b'(a'a')\bar{b}'$. All 8 $A_g$ modes are successfully observed. The eigen-frequency is extracted at the maximum intensity of the different Raman bands. The eigen-frequencies are summarized in Table~\ref{table:ModeAssignm} for both polarizations. As expected, the eigen-frequencies of the individual Raman bands compares well between the two polarization conditions. For the $A_g^7$ mode, however, a significant difference in the eigen-frequency is observed from one polarization to the other. The origin for this difference is phonon-phonon coupling between the $A_g^7$ and $A_g^8$ mode which is discussed in the main paper.

\begin{table}[htp]
\centering
\caption{Raman eigen-frequencies for both polarization geometries assigned to the active Raman modes with $A_g$ symmetry. The eigen-frequencies are given in units of \SI{}{\wn}.}
\begin{tabular}{c c c} 
 \hline
 Mode & $\omega_{(cc)}$ & $\omega_{(a'a')}$\\ [0.5ex] 
 \hline
 $A_g^1$  & 65 & 65\\
 $A_g^2$  & 130 & 130 \\ 
 $A_g^3$   & 216 & 214\\
 $A_g^4$   & 328 & 328\\ 
 $A_g^5$   & 368 & 370\\ 
 $A_g^6$   & 385 & -\\
 $A_g^7$   & 701 & 711\\ 
 $A_g^8$   & 830 & 830\\ 
 \hline
\end{tabular}
\label{table:ModeAssignm}
\end{table}

\clearpage

\section*{Evidence for phonon-phonon mode coupling - Polarization dependent temperature trend of the $A_g^7$ mode (Lorentzian fitting)}

\begin{figure}[h!]
\centering
 \includegraphics[width=\linewidth]{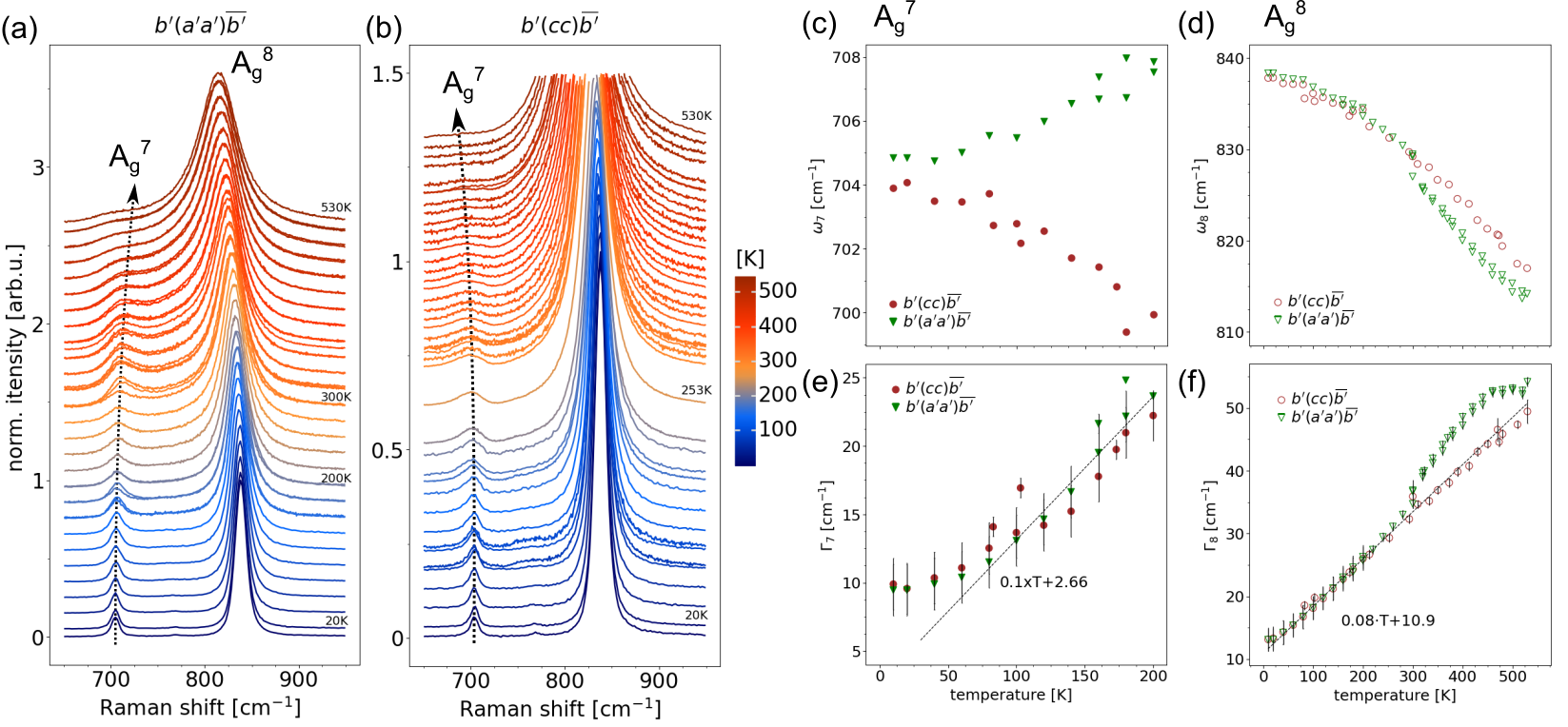}
\caption{High frequency $A_g^7$ and $A_g^8$ Raman mode of monoclinic \BVO measured for the two polarization geometries (a) $b'(a'a')\bar{b}'$ and (b)  $b'(cc)\bar{b}'$ from \SI{10}{\K} up to \SI{530}{\K} with \SI{633}{\nm} laser excitation. (c) and (d) show the eigen-frequencies $\omega_7$, $\omega_8$ and (e) and (f) the damping coefficient $\Gamma_7$ and $\Gamma_8$ obtained from the Lorentzian fitting. The damping coefficients increase nearly linear with increasing temperatures. The linear fits are shown with black lines and the fitting equations are provides in the plots.}
\label{fig:Lorentzian_HF}
\end{figure}

Figs.~\ref{fig:Lorentzian_HF}~(a) and (b) show the high frequency $A_g^7$ and $A_g^8$ Raman modes measured as a function of temperature for the $b'(a'a')\bar{b}'$ and the $b'(cc)\bar{b}'$ polarization geometry, respectively. The $A_g^8$ mode clearly shifts to lower frequencies with increasing temperatures for both polarization geometries. The $A_g^7$ mode shift to higher frequencies for $b'(a'a')\bar{b}'$ whereas the temperature trend is opposite for $b'(cc)\bar{b}'$. The opposite temperature trend of the $A_g^7$ is surprising. In first approximation, there is no strong argument for the frequency position to be different between the two polarization geometries. This is a first indication that the $A_g^7$ strongly couples to the $A_g^8$ mode. Secondly, the line-shape of the $A_g^7$ is strongly asymmetric at temperatures above room temperature. 

In a first attempt, $A_g^7$ and $A_g^8$ modes have been fitted with two Lorentzian functions.  Because of the asymmetric line-shape of the $A_g^7$ mode at high temperatures, the Lorentzian function was used only from \SI{10}{\K} to \SI{200}{\K} for this mode. For the $A_g^8$ mode the Lorentzian function seem to provide a good fitting result up to the phase transition temperature. Figs.~\ref{fig:Lorentzian_HF}~(c) and (d) show the eigen-frequencies of the $A_g^7$ and $A_g^8$ mode as a function of temperature and polarization geometry, respectively. The eigen-frequency of the $A_g^8$ mode decreases with increasing temperature for both scattering geometries whereas the eigen-frequency of the $A_g^7$ has opposite temperature trend depending on the polarization geometry. The damping coefficients of both modes increase with temperatures as expected, see Figs.~\ref{fig:Lorentzian_HF}~(e) and (f). For temperatures above \SI{100}{\K}, the increase is nearly linear for both phonon modes and both scattering geometries. The linear fits are shown with black lines and the fitting equations are given in the corresponding plots. The linear temperature trend was used in the main paper to physical constrain the damping $\Gamma_{7}$ and $\Gamma_{8}$ of the $A_g^7$ and $A_g^8$ for the coupled oscillators model.

\clearpage

\section*{Coupled oscillators model - Oscillator strengths}

\begin{figure}[h!]
\centering
 \includegraphics[width=0.5\linewidth]{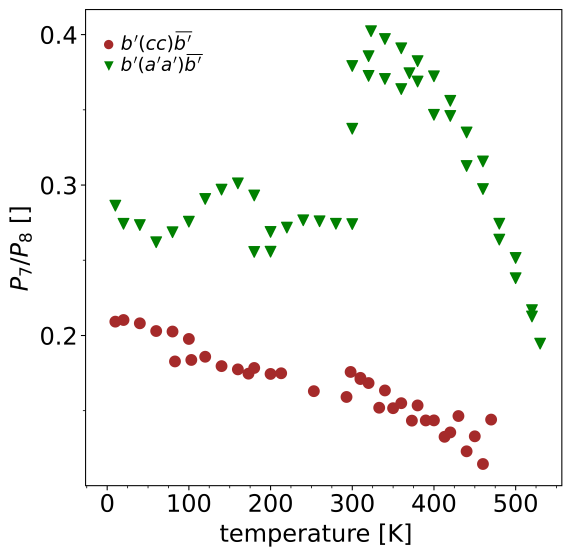}
\caption{Oscillator strengths ratio $P_7/P_8$ as a function of temperature for the polarization condition $b'(cc)\bar{b'}$ and $b'(a'a')\bar{b'}$ obtained from the coupled oscillators model.}
\label{fig:CDHO_Amp}
\end{figure}

Fig.~\ref{fig:CDHO_Amp} shows the ratio of the oscillator strengths $P_7/P_8$ as a function of the temperature for the polarization condition $b'(cc)\bar{b}'$ and $b'(a'a')\bar{b}'$. The ratio of the oscillator strengths $P_7/P_8$ decreases while approaching the phase transition for both polarization conditions. That is consistent with the observation on the raw data where the intensity of the $A_g^7$ is extremely weak at high temperatures at which the $A_g^8$ mode dominates.

